\documentclass[aps, prx,twocolumn,longbibliography,superscriptaddress,amsmath,amssymb,floatfix]{revtex4}
\usepackage[latin9]{inputenc}
\usepackage{amssymb}
\usepackage{graphicx}
\usepackage{subfigure}
\usepackage{amsmath}
\usepackage{color}
\usepackage{mathrsfs}
\usepackage{float}
\usepackage{indentfirst}
\usepackage{mathrsfs}
\usepackage{float}
\usepackage{indentfirst}
\usepackage{textcomp}
\usepackage{comment}
\usepackage{mathtools}
\usepackage{natbib,hyperref}
\usepackage{soul}

\begin{document}

\title{Reexamining doped two-legged Hubbard ladders}

\author{Yang Shen}
\affiliation{Key Laboratory of Artificial Structures and Quantum Control (Ministry of Education),  School of Physics and Astronomy, Shanghai Jiao Tong University, Shanghai 200240, China}

\author{Guang-Ming Zhang}
\affiliation{State Key Laboratory of Low-Dimensional Quantum Physics and Department of Physics, Tsinghua University, Beijing 100084, China}
\affiliation{Frontier Science Center for Quantum Information, Beijing 100084, China}

\author{Mingpu Qin} \thanks{qinmingpu@sjtu.edu.cn}
\affiliation{Key Laboratory of Artificial Structures and Quantum Control (Ministry of Education),  School of Physics and Astronomy, Shanghai Jiao Tong University, Shanghai 200240, China}
\affiliation{Hefei National Laboratory, University of Science and Technology of China, Hefei 230088, China}

\begin{abstract}
We revisit the ground state of the Hubbard model on 2-legged ladders in this work. We perform DMRG calculation on large system sizes with large kept states and perform extrapolation of DMRG results with truncation errors in the converged region. We find the superconducting correlation exponent $K_{sc}$ extracted from the pair-pair correlation is very sensitive to the position of the reference bond, reflecting a huge boundary effect on it. By systematically removing the effects from boundary conditions, finite sizes, and truncation errors in DMRG, we obtain the most accurate value of $K_{sc}$ and $K_\rho$ so far with DMRG. With these exponents, we confirm that the 2-legged Hubbard model is in the Luther-Emery liquid phase with $K_{sc} \cdot K_\rho = 1$ from tiny doping near half-filling to $1/8$ hole doping. When the doping is increased to $\delta \gtrapprox 1/6$, the behaviors of charge, pairing, and spin correlations don't change qualitatively, but the relationship $K_{sc} \cdot K_\rho = 1$ is likely to be violated. With the further increase of the doping to $\delta = 1/3$, the quasi long-ranged charge correlation turns to a true long-ranged charge order and the spin gap is closed, while the pair-pair correlation still decays algebraically. Our work provides a standard way to analyze the correlation functions when studying systems with open boundary conditions.
\end{abstract}

\maketitle

\section{Introduction}
The Hubbard model \cite{Hubbard,doi:10.1146/annurev-conmatphys-031620-102024,qin2022hubbard} is the prototype lattice model of interacting Fermions and plays a paradigmatic role in correlated electron physics \cite{dagotto1994correlated}. The connection of the two-dimensional Hubbard model and the related t-J model to cuprate superconductors has been debated \cite{PhysRevLett.88.117001,PhysRevLett.110.216405,PhysRevB.98.205132,PhysRevB.100.195141,PhysRevX.10.031016,PhysRevB.102.041106,doi:10.1073/pnas.2109978118,PhysRevLett.127.097003,PhysRevLett.127.097002,2023arXiv230308376X} ever since the discovery of the first cuprate superconductor \cite{bednorz1986possible}. Because there is no rigorous analytic solution to the Hubbard model beyond one dimension \cite{PhysRevLett.20.1445}, most studies of it rely on numerical methods \cite{PhysRevX.5.041041}.

Quasi-one-dimensional systems, such as 2-legged Hubbard ladders, were extensively studied in the past \cite{PhysRevB.53.251,PhysRevB.53.12133,RevModPhys.77.259,PhysRevB.68.115104,he2023terminable,zhou2023robust}. On one hand, the correlations in the 2-legged Hubbard model can be taken as the precursor to the possible superconducting and charge density wave (CDW) instabilities in the two-dimensional systems. On the other hand, 2-legged systems can be accurately solved by Density Matrix Renormalization Group (DMRG) \cite{PhysRevLett.69.2863,PhysRevB.48.10345,SCHOLLWOCK201196}.
Early analytic and numerical works \cite{PhysRevB.53.251,PhysRevB.53.12133,RevModPhys.77.259} demonstrate that the slightly doped ladders can be categorized into the Luther-Emery liquid phase \cite{PhysRevLett.33.589,PhysRevLett.45.1358}, which promises a single gapless charge mode and a gapped spin mode (labeled as C1S0 in the literature \cite{PhysRevB.53.12133, PhysRevB.92.195139, lu2022ground}).


The Luther-Emery liquid phase is characterized by the algebraic decay of both charge (with exponent $K_\rho$) and pair-pair correlations (with exponent $K_{sc}$).  Moreover, the exponents are related and satisfy the relationship $K_{sc} \cdot K_\rho = 1$ \cite{PhysRevLett.33.589,PhysRevB.92.195139,lu2022ground}.
Numerically, whether the relationship $K_{sc}  \cdot K_\rho = 1$ holds was debated with both positive \cite{PhysRevB.92.195139} and negative \cite{PhysRevB.56.7162,NOACK1996281} results. $K_\rho$ can be usually extracted from the Friedel oscillations \cite{PhysRevB.65.165122} in a system with open boundaries. The difficulty to obtain a reliable $K_{sc}$ lies in the subtlety in the extraction of the exponent $K_{sc}$ from the pair-pair correlation function, which usually oscillates with distance under the influence of the modulation of charge density \cite{PhysRevB.56.7162,NOACK1996281,PhysRevB.92.195139}.
In the strong-coupling limit of the $t-J$ model with hole doping $\delta \xrightarrow{} 0$, bosonization calculation yields $K_{sc} = 1/2$ \cite{PhysRevB.59.R2471,PhysRevB.63.195106}. Previous DMRG calculations found the decay of the superconducting correlation function to be much faster than $1/K_\rho$ at small doping, hence violates the relationship $K_{sc} \cdot K_\rho = 1$ \cite{PhysRevB.56.7162,NOACK1996281}. But later results \cite{PhysRevB.92.195139} found the relationship of the exponents holds. Therefore, it is natural to ask to what degree, $e.g.$, doping levels, the relationship of the exponents holds in the 2-legged Hubbard model.

\begin{table}[t]
    \caption{Summary of the extracted exponents and correlation lengths. $K_\rho$ and $K_{sc}$ are the exponents for the algebraic decay of charge density and pair-pair correlations. $\xi_s$ and $\xi_G$ are the correlation lengths for spin correlation (local magnetization) and single-particle Green's function. Quantities not listed in the table are ill-defined.}
    \includegraphics[width=0.45\textwidth]{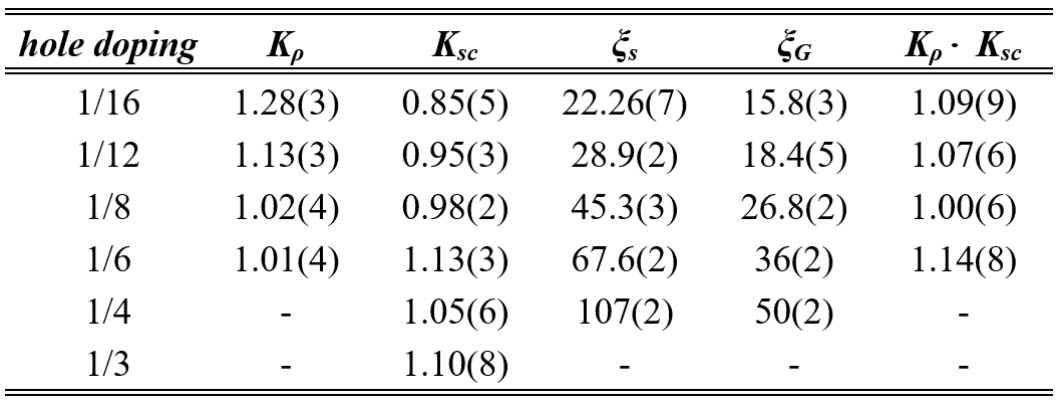}
    \label{Summary}
\end{table}

In this work, we revisit the ground state of the 2-legged Hubbard model. We study dopings ranging from $1/16$ to as large as $1/3$, trying to figure out the boundary of the Luther-Emery liquid phase by examining the behaviors of charge, spin, and pair-pair correlations. We focus on the issue whether the relationship $K_{sc} \cdot K_\rho = 1$ holds. 
We employ DMRG in this work which can provide extremely accurate results for ladder systems nowadays with the increase of computational power. We push the bond dimension to as large as $9500$ (with truncation error at the order of $10^{-7}$) to achieve an unprecedented accuracy for the Hubbard model on ladders \cite{PhysRevB.56.7162,NOACK1996281,PhysRevB.92.195139}. More importantly, we study large system sizes and make sure the extracted exponents are free from finite size effect. We also vary the position of the reference bond in the calculation of pair-pair correlation functions to get rid of the effect of open boundaries, which turn out to have a huge impact on the extracted exponent $K_{sc}$ (the error caused by the open boundaries could be as large as $50\%$ as will be shown in the discussion of results). With these careful treatment of effects of boundary conditions, finite sizes and truncation error in DMRG calculations, we give the most accurate values of $K_\rho$ and $K_{sc}$ so far with which we can resolve the previous controversy on whether the relationship $K_{sc} \cdot K_\rho = 1$ holds and determine the precise phase boundary of the Luther-Emery liquid phase.

\begin{figure}[t]
    \includegraphics[width=0.45\textwidth]{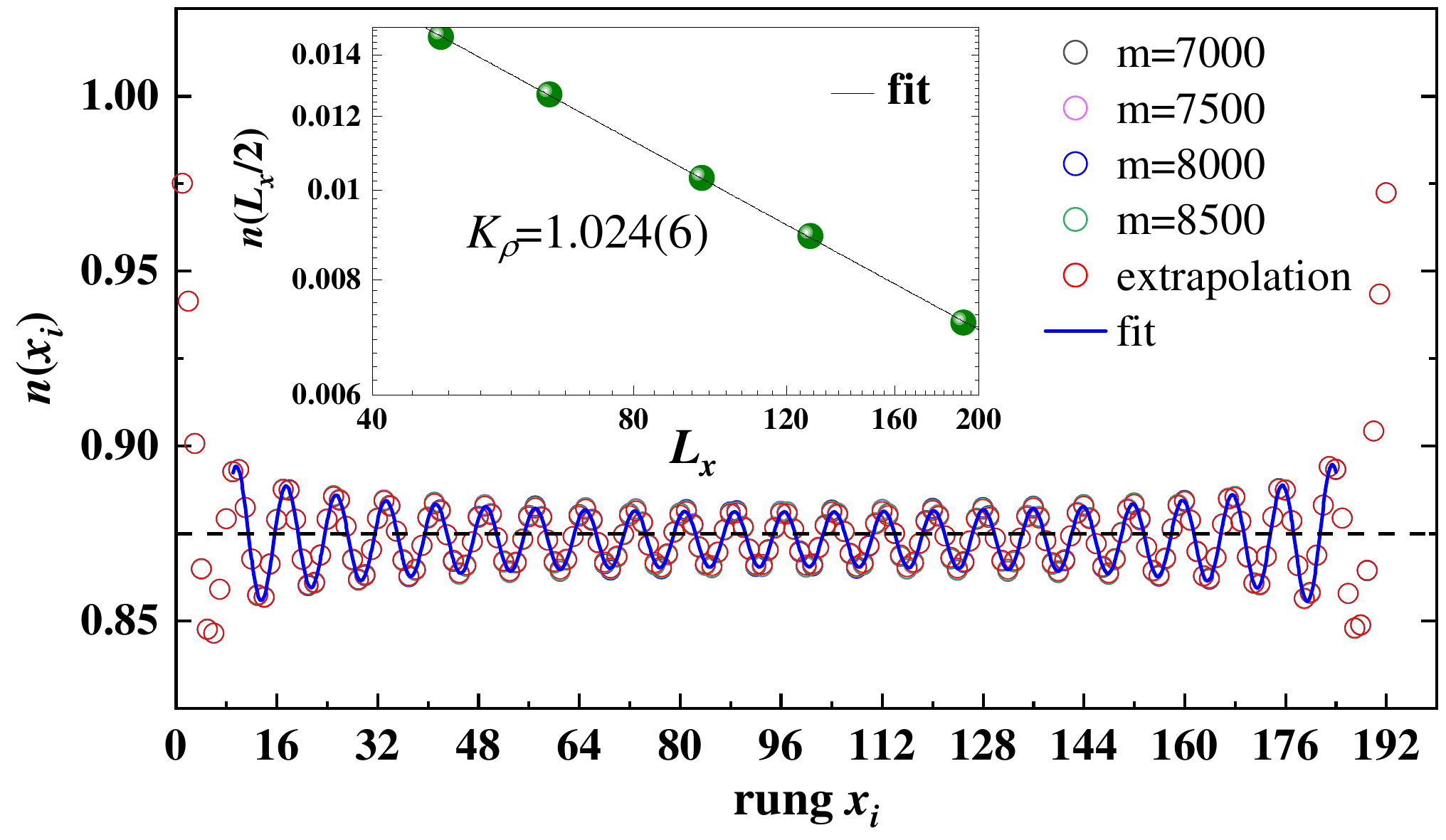}
    \caption{Density profile at 1/8 doping. The local rung density $n(x)$ for different kept states $m$ and the extrapolated with truncated errors results are shown. The length of the system is $L_x =192$. The solid line is the fitting curve using Eq.~\eqref{charge_oscillation} to extract $K_\rho$. The dashed horizontal line represents the averaged electron density. The inset is the finite-size scaling of $\delta n(L_x/2)$ as a function of the system size $L_x$ using Eq.~(\ref{finite_scale}).}
    \label{ElecOneeighthL192}
\end{figure}

The main results are summarized in Table ~\ref{Summary}. We focus on the strongly correlated region with $U/t = 8$. From tiny doping near half-filling to $\delta = 1/8$, 2-legged Hubbard model is found to be in the Luther-Emery liquid phase with algebraic decay of pair-pair and charge correlations and exponential decay of spin correlation. The exponents of charge and pairing correlations also satisfy the relationship $K_{sc} \cdot K_\rho =1$. In this region, $K_{sc} < K_\rho$ which means superconducting is the dominant correlation. The increase of the hole doping tends to suppress the pairing correlations, and consequently enhances the charge correlation. By increasing the doping to $1/6$, the characterization of charge, pairing, and spin correlations don't change qualitatively, but the exponents of superconducting and charge correlations seem to violate the $K_{sc} \cdot K_\rho = 1$ relationship. The system also switches to a charge correlation dominating phase with $ K_\rho < K_{sc}$.  
{By further increasing doping level, $i.e.$, for $\delta=1/4$ and $1/3$, $K_\rho$ is ill-defined and the charge correlation is likely to be long-ranged. The spin gap is closed at $1/3$ doping but the algebraic decay of pair-pair correlation functions remains.

We notice that $K_\rho$ in Table~\ref{Summary} are consistent with the recently results obtained from infinite system matrix product states \cite{2023arXiv230300663E}.

\begin{figure}[t]
    \includegraphics[width=0.45\textwidth]{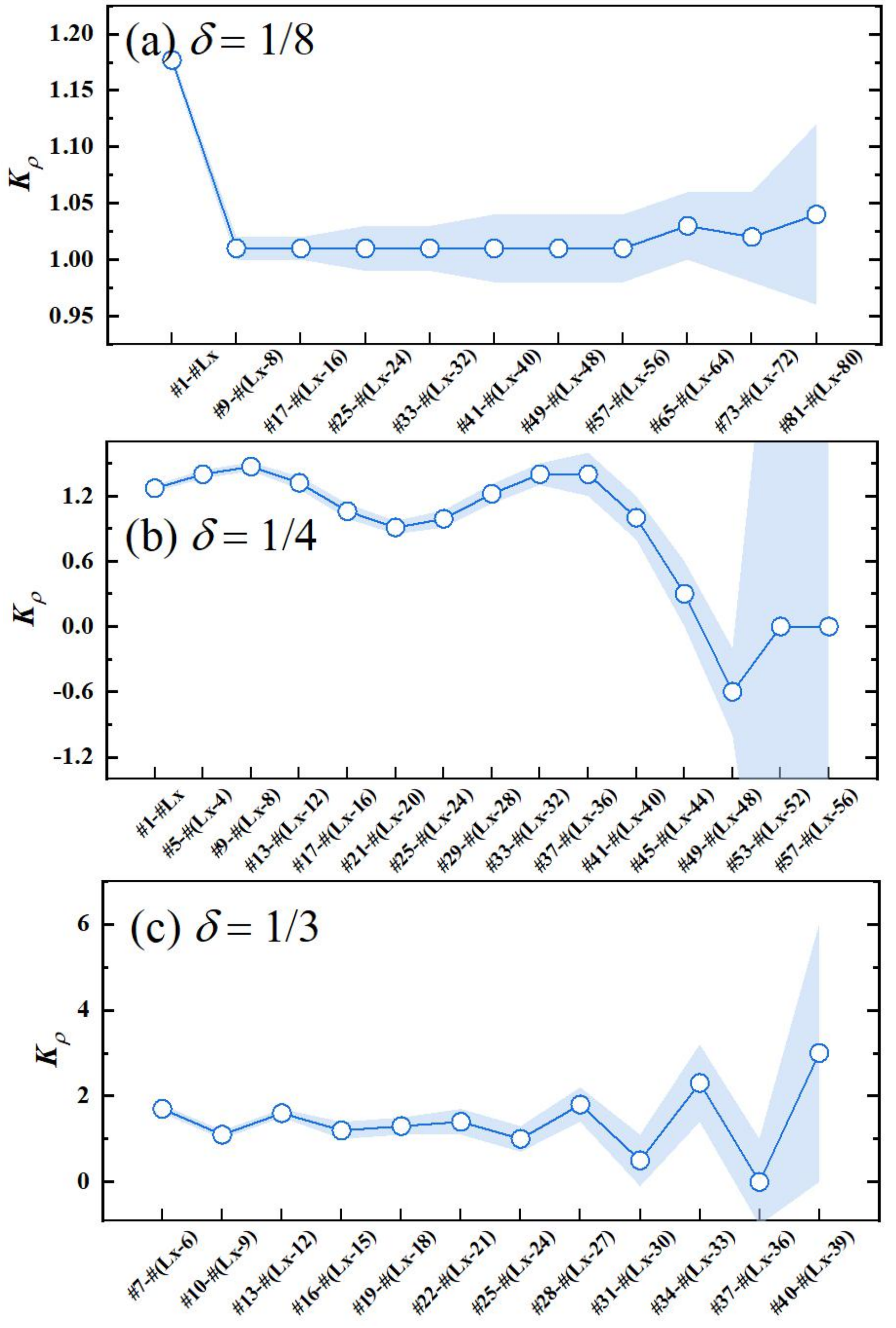}
    \caption{The dependence of the extracted parameters $K_\rho$ (using Eq.~\eqref{charge_oscillation}) on the range of sites used for (a) $\delta=1/8$ ($L_x$ = 192), (b) $\delta=1/4$ ($L_x$ = 128) and (c) $\delta=1/3$ ($L_x$ = 96). The shaded region around the curve shows the error bar. The fitted values of $K_\rho$, $n_0$ and $R_2$ are also listed in Tables.~\ref{ElecOscillationFit-oneeighthL192},~\ref{ElecOscillationFit_onefourthL128} and ~\ref{ElecOscillationFit_onethirdL96}.}
    \label{Fig2}
\end{figure}

\begin{table}[t]
    \caption{The dependence of the extracted parameters $K_\rho$ and $n_0$ (using Eq.~\eqref{charge_oscillation}) on the range of sites used ($\delta=1/8$, $L_x$ = 192). Also see a plot of $K_\rho$ versus the range of site used in Fig.~\ref{Fig2}(a).}
    \includegraphics[width=0.3\textwidth]{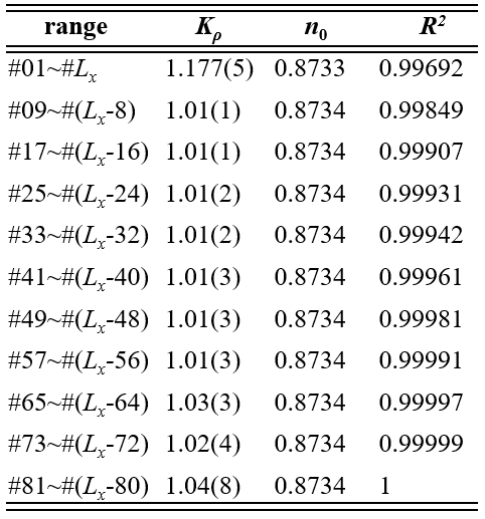}
    \label{ElecOscillationFit-oneeighthL192}
\end{table}

The remaining of the paper is organized as follows. In Sec.~\ref{sec_METH} we introduce the Hubbard model and discuss the calculation details. We then provide results of different doping levels in Sec.~\ref{sec_RES}. We show the long-distance behaviors of the pairing, charge, spin and single-particle correlations. We extract the exponents and correlation lengths from them by carefully analyzing the effects from boundary conditions, finite sizes, and truncation error in DMRG calculations. We finally summarize this work in Sec.~\ref{sec_con}.

\section{MODEL AND METHODOLOGY}
\label{sec_METH}
The Hamiltonian of the Hubbard model is
\begin{equation}
\begin{aligned}
\hat{H} =-\sum_{\langle i,j \rangle, \sigma} t_{i j}\left(\hat{c}_{i \sigma}^{\dagger} \hat{c}_{j \sigma}+h.c.\right)+U \sum_{i } \hat{n}_{i \uparrow} \hat{n}_{i \downarrow}
\end{aligned}
\label{Ham}
\end{equation}
where $\hat{c}_{i \sigma}^{\dagger}(\hat{c}_{j \sigma})$  creates (annihilates) an electron on site $i=(x_i,y_i)$ with spin $\sigma$, and $U$ represents the on-site Coulomb interaction. We only consider nearest hopping $t$ and set $U = 8t$. $\hat{n}_{i}=\sum_{\sigma} \hat{c}_{i \sigma}^{\dagger} \hat{c}_{i \sigma}$ is the electron number operator.

We focus on ladders with width $L_y \equiv 2$ and length $L_x$, so there are totally $N=L_x \times L_y$ lattice sites. The average hole concentration away from half-filling is defined as $\delta = N_h/N$ with $N_h=\sum_i(1-\hat{n}_i)$. We apply a one-site pinning field (an addition term $h_m\cdot \hat{S}_{0z}$  with $h_m = 0.5$ is added to the Hamiltonian in Eq.~(\ref{Ham})) at the left edge of the ladder, which allows us to detect the local magnetization $\langle \hat{S}_{i}^z \rangle  (=1/2 \times (\langle \hat{n}_{i \uparrow} \rangle - \langle \hat{n}_{i \downarrow}\rangle))$ instead of the more demanding correlation functions. We also measure the $d$-wave pair-pair correlation function defined as $D(r) = \langle \hat{\Delta}_i^{\dagger}\hat{\Delta}_{i+r}\rangle$ with $\hat{\Delta}_i^{\dagger}=\hat{c}_{(i,1),\uparrow}^{\dagger}\hat{c}_{(i,2),\downarrow}^{\dagger}-\hat{c}_{(i,1),\downarrow}^{\dagger}\hat{c}_{(i,2),\uparrow}^{\dagger}$ creating a singlet on the rungs \cite{PhysRevB.92.195139,shen2022comparative}. Other physical observables are defined wherein they are firstly mentioned.

We employ DMRG method which can provide extremely accurate results for ladder systems. We push the bond dimension in DMRG calculation up to $m=9500$ with typical truncation errors $\epsilon$ at the order of  $10^{-7}$. Extrapolations with truncation errors $\epsilon$ are also performed to remove the tiny residual truncation errors. We study the system as long as $192$ sites to get rid of the finite size effect. More importantly, when calculating the pair-pair correlation function, we vary the position of reference bond to ensure the effect of boundary conditions is absent. It turns out the boundary conditions have a huge effect (could cause an error as large as $50\%$) on the extracted value of $K_{sc}$. With the careful treatment of the possible errors, we provide reliable results on the charge density, pair-pair correlations and local magnetization.  

\section{RESULTS AND DISCUSSION}
\label{sec_RES}

\subsection{1/8 doping level}

\begin{figure}[t]
    \includegraphics[width=0.4\textwidth]{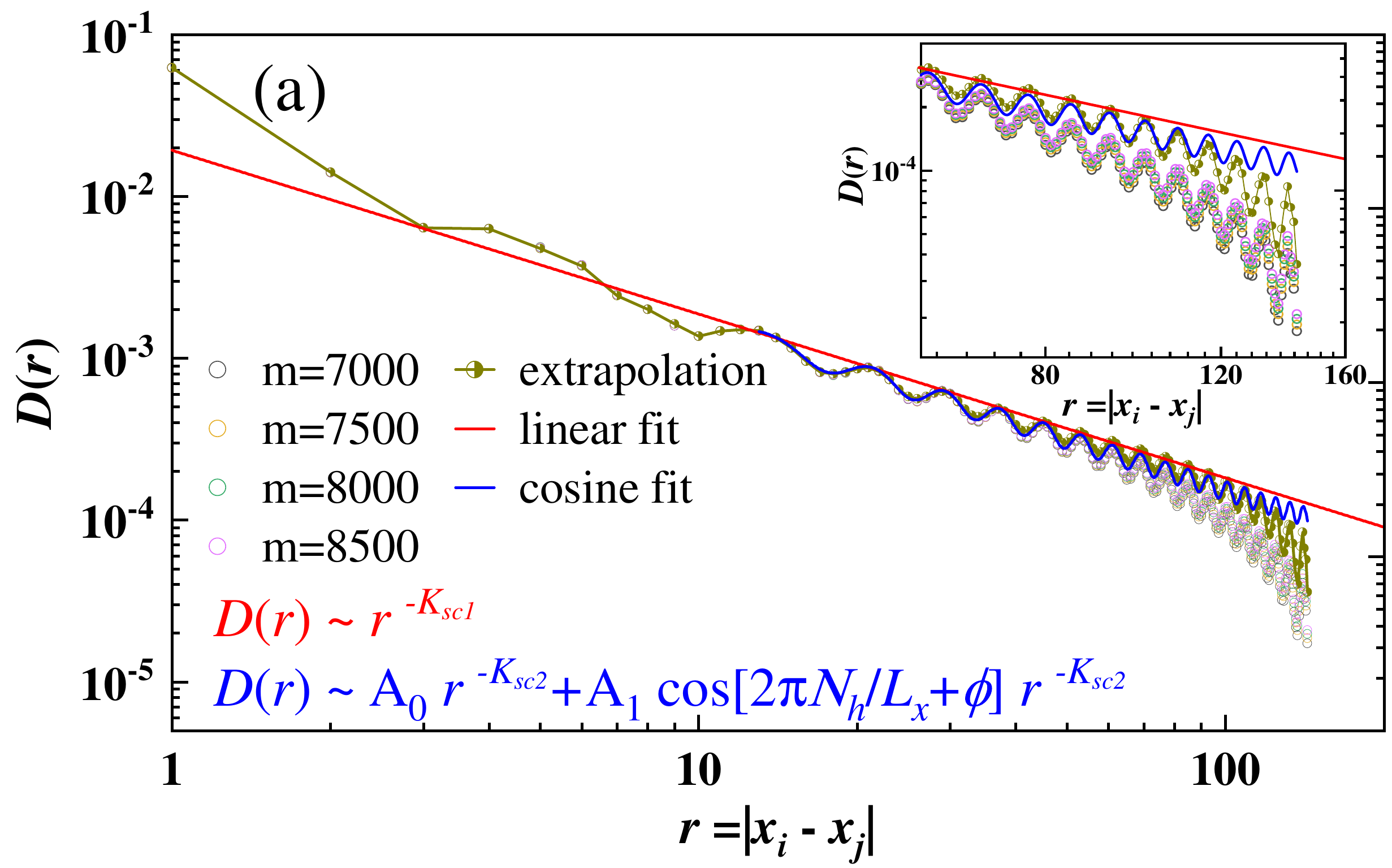}
    \includegraphics[width=0.4\textwidth]{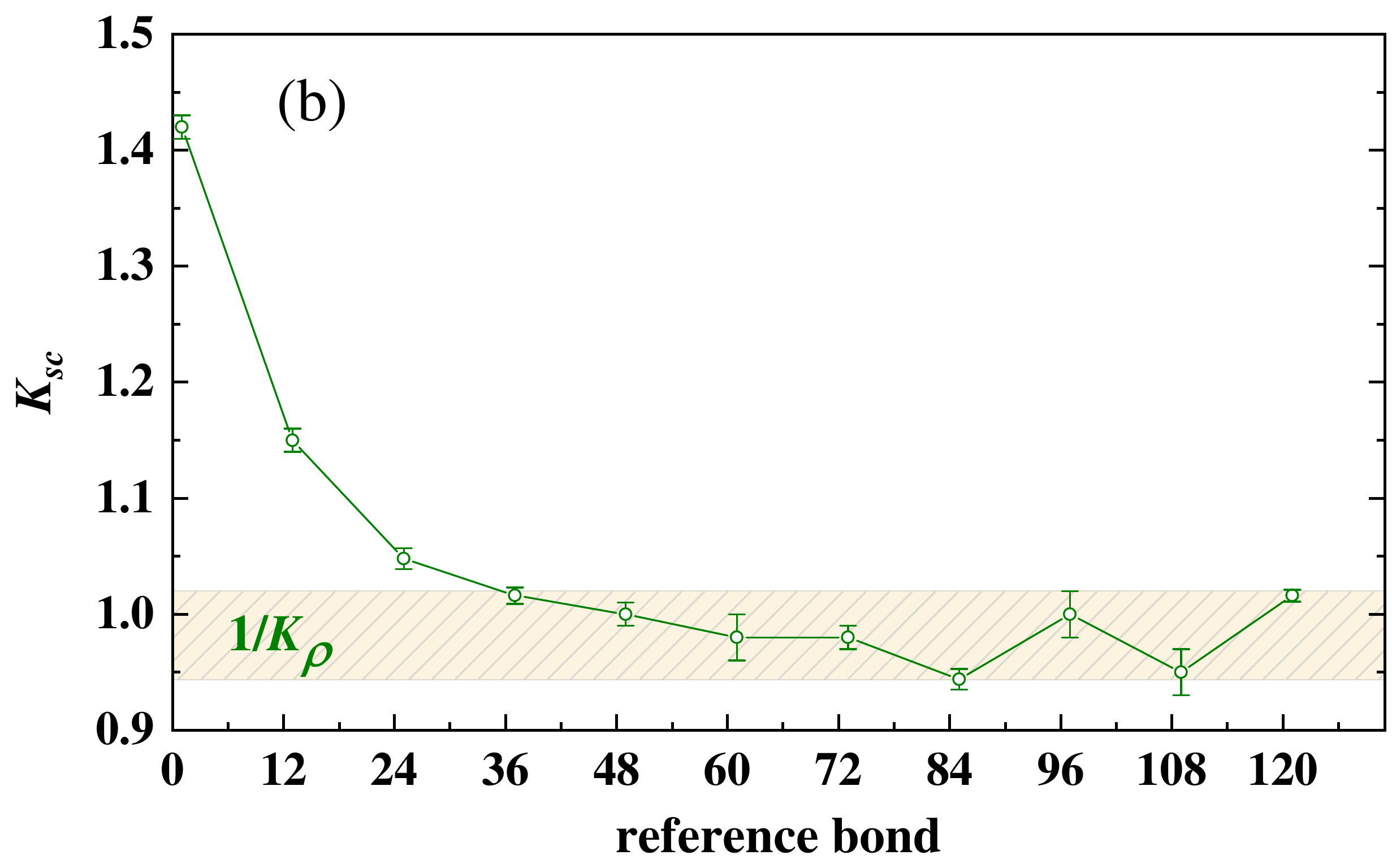}
    \caption{Pair-pair correlation and the extracted exponents at $1/8$ doping. (a) Pair-pair correlation function $D(r)$ for different numbers of kept states $m$ and the extrapolated result. The length of the system is $L_x = 192$. {The reference bond is set at the 49th vertical bond.}  The red and blue lines denote linear and cosine fits defined in the main text, respectively. Only peaked values are used in the linear fit. The inset zooms in the data scale. (b) $K_{sc}$ as a function of the reference bond. Here we only show the results of $K_{sc1}$ from linear fits ($K_{sc2}$ has similar value). 
    The shaded area represents the value of $1/K_{\rho}$ with error bar. }
    \label{PairOneeighth}
\end{figure}

\begin{figure}[t]
    \includegraphics[width=0.4\textwidth]{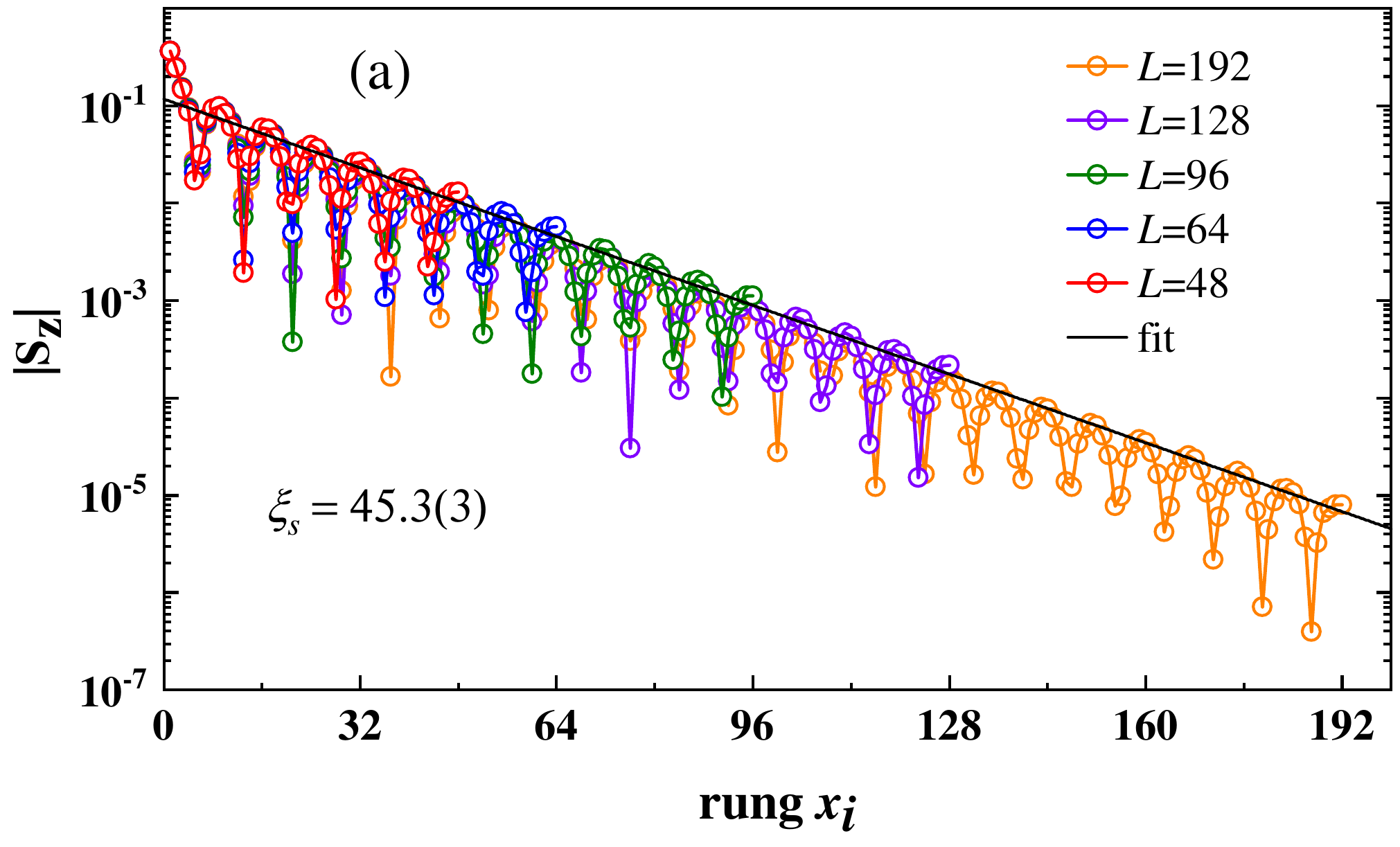}
    \includegraphics[width=0.4\textwidth]{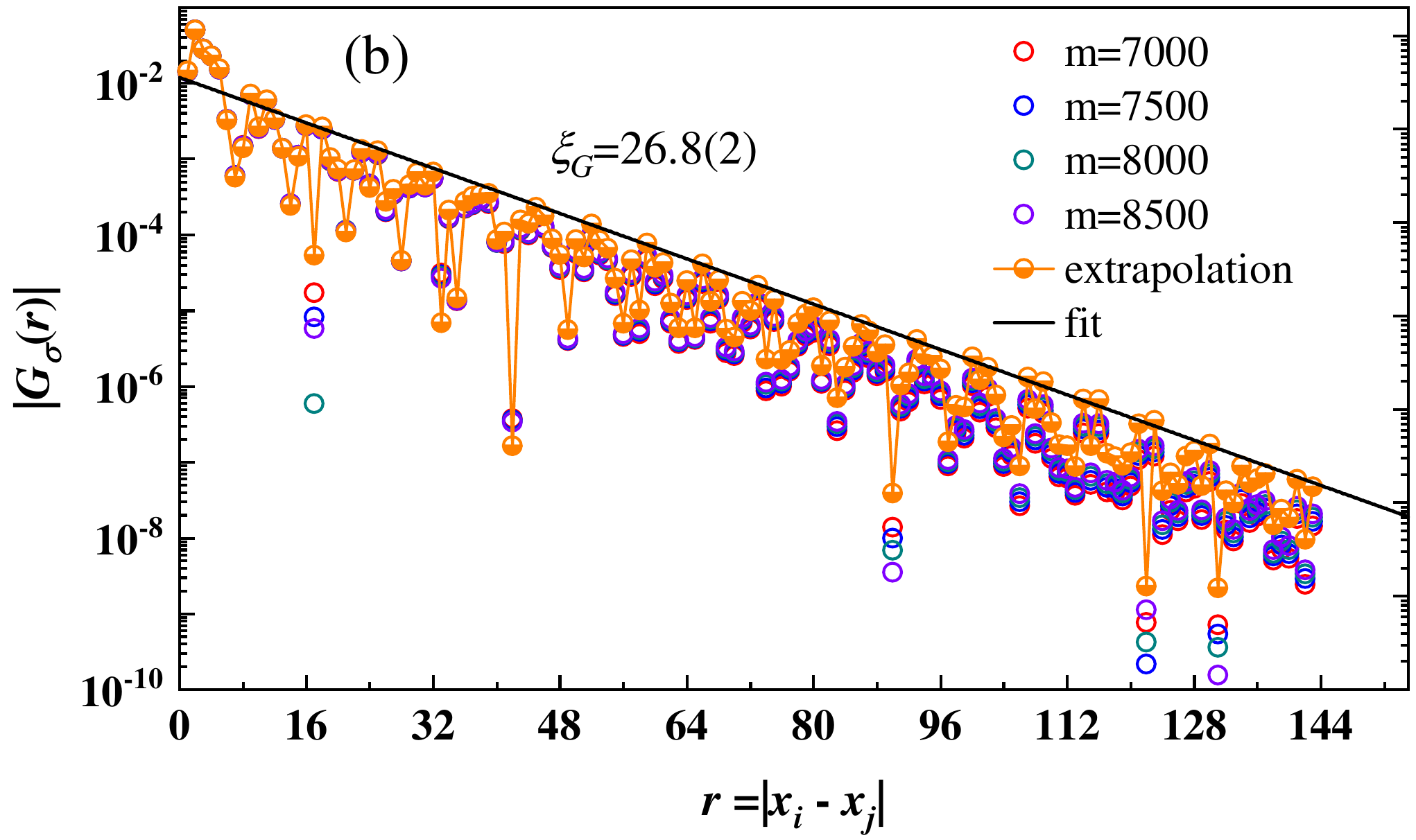}
    \caption{Local magnetization and single-particle Green's function at $1/8$ doping. 
    (a) Absolute values of the local magnetization with different lengths. We only show the local magnetization on one leg. $S_z$ on different legs are identical in absolute value but with opposite sign. This is also true for other plots of $S_z$. Only the results from the extrapolation with truncation errors are shown. The solid line denotes the exponential fit using $\left| \mathbf{\textit{S}}_\mathbf{\textit{z}}\right|  \propto \mathrm{e}^{-x_i / \xi_{\mathrm{s}}}$ with $\xi_{\mathrm{s}}=45.3(3)$  for $L_x=192$. (b) Single-particle Green's function with different numbers of kept states $m$ and the extrapolated result for $L_x$ = 192. The reference site is set at (49, 2). Only peaked values are used in the fits for both the local magnetization and the single-particle Green's function.}
\label{OneeighthSz_Gf}
\end{figure}

We start with the 1/8 doping case, which draws dramatic interests because of the intertwined spin, charge, and pairing correlation near $1/8$ doping in cuprates \cite{tranquada1995evidence,doi:10.1073/pnas.96.16.8814}. Fig.~\ref{ElecOneeighthL192} shows the charge density profiles of a ladder with length $192$, defined as the rung density $n(x_i)=\sum_{y=1}^{L_{y}}\left\langle\hat{n}_{i}(x,y)\right\rangle / L_{y}$ on the ladders. The charge density from DMRG with kept states $m = 7000$ to $m = 8500$ and the result from an extrapolation with truncation error are all on top of each other, indicating the DMRG results are well converged. For the 1/8 hole doping case in Fig.~\ref{ElecOneeighthL192}, the charge distribution $n(x)$ forms a CDW pattern with a wavelength $\lambda_c=1/\delta=8$. The spatial decay of the CDW correlation at long distances can be described by a power law with Luttinger exponent $K_\rho$. Previous works \cite{PhysRevB.65.165122,PhysRevB.92.195139} show that $K_\rho$ can be obtained by fitting the Friedel oscillations induced by the boundaries of the ladder
\begin{equation}
n(x) \approx A \frac{\cos \left(2 \pi N_{h} x / L_x+\phi_{1}\right)}{\left[L_x \sin \left(\pi x / L_x+\phi_{2}\right)\right]^{K_{\rho} / 2}}+n_{0}
\label{charge_oscillation}
\end{equation}
where $A$ is an amplitude, $\phi_1$ and $\phi_2$ are phase shifts, and $n_0=1-\delta$ is the averaged charge density (we set $n_0$ as a parameter to be determined in the fit \cite{PhysRevB.92.195139} ). 
 To get rid of the boundary effect, we fit the density by gradually excluding data near the boundaries, and the results are shown in Fig.~\ref{Fig2}(a) and Table~\ref{ElecOscillationFit-oneeighthL192}. We can find an obvious boundary effect on the extracted exponent $K_\rho$. After excluding about a dozen of data near the boundary, the extracted $K_\rho$ converges fast to a value about $1.02(4)$ and the $R^2$ of the fit is very close to 1 (with a deviation in the order of 0.0001). We also study ladders with lengths 48, 64, 96, and 128 and find the extracted values of $K_\rho$ is converged. For comparison, we can alternatively obtain the parameter $K_\rho$ from the finite-size scaling, because the density at the center of the system scales as \cite{PhysRevB.92.195139}
\begin{equation}
\delta n(L_x/2)=n(L_x/2)-n_{0} \sim {L_x}^{-K_{\rho} / 2}
\label{finite_scale}
\end{equation}
The advantage of this scheme is that $n(L_x/2)$ is least affected by the boundary effect. The finite-size fit is shown in the inset of Fig.~\ref{ElecOneeighthL192}. 
 To calculate $\delta n(L_x/2)$, we use the obtained $n_0$ from the least-square fit of the charge density profiles using Eq.~\eqref{charge_oscillation} \cite{PhysRevB.92.195139}. 
Considering the values from the two fitting procedures, we give an estimation of $K_\rho$ at $1/8$ as $K_\rho \approx 1.02(4)$. The values of $K_\rho$ for other dopings are estimated in the same way.


\begin{figure}[t]
    \includegraphics[width=0.4\textwidth]{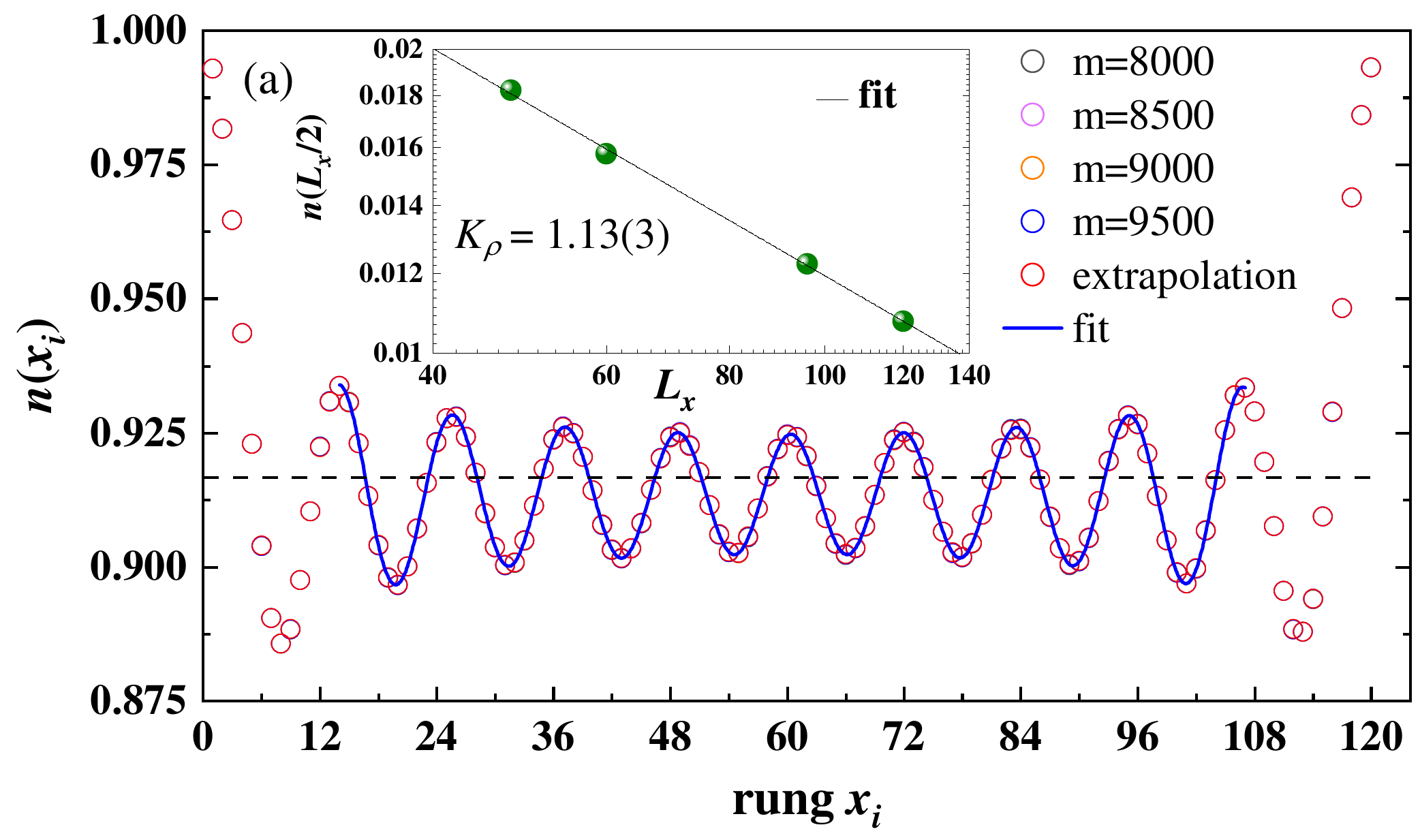}
    \includegraphics[width=0.4\textwidth]{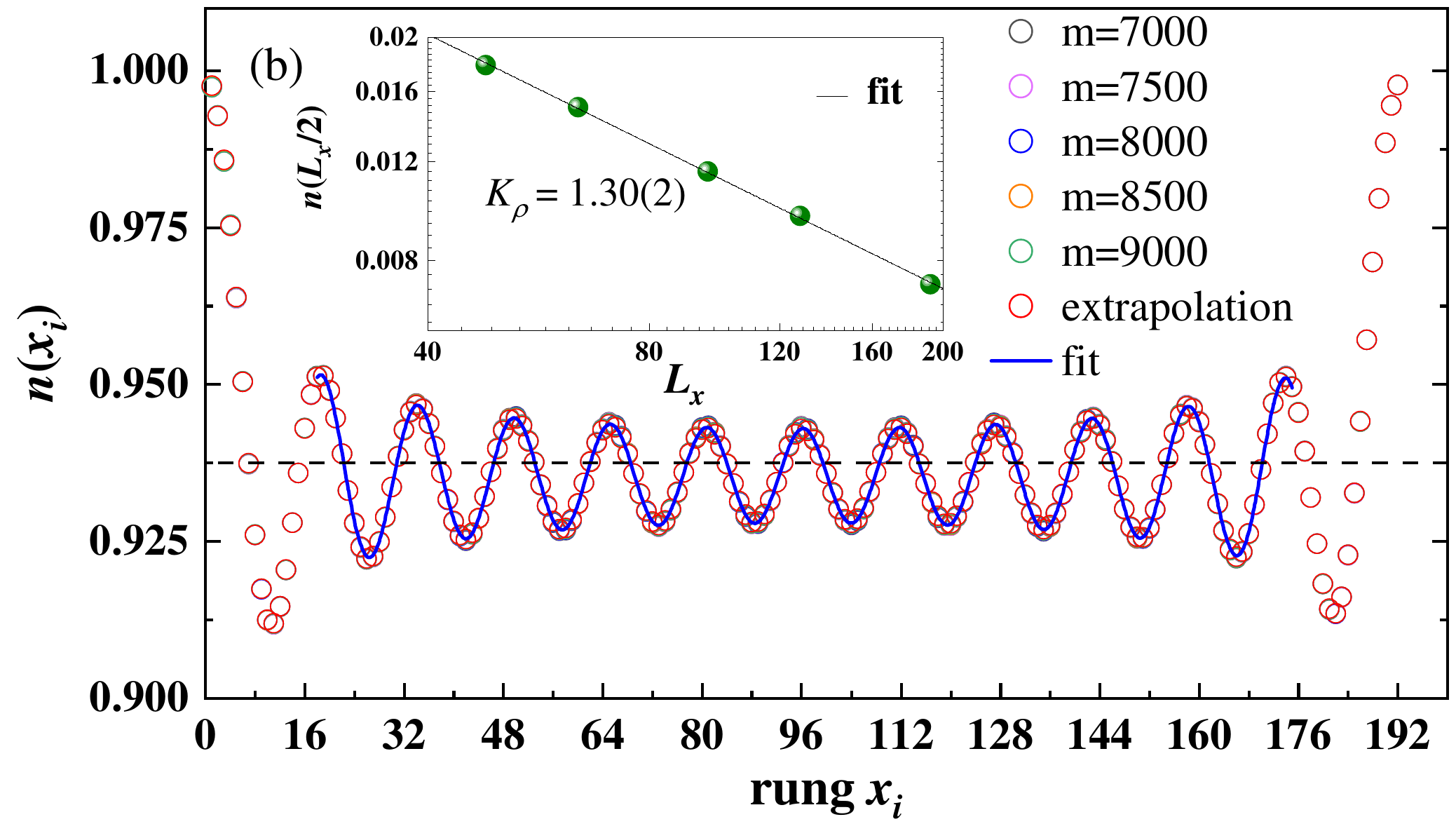}
    \caption{Density profiles at (a) $1/12$ and (b) $1/16$ dopings. The local rung density $n(x)$ for different kept states $m$ and the extrapolated with truncated errors results are shown. The lengths of the systems are $L_x =120$ for $1/12$ doping and $L_x =192$ for $1/16$ doping, respectively. The solid lines are the fitting curves using Eq.~\eqref{charge_oscillation} to extract $K_\rho$. The dashed horizontal lines represent the averaged electron density. The insets of (a) and (b) are finite-size scaling of $\delta n(L_x/2)$ as a function of the system size $L_x$ using Eq.~(\ref{finite_scale}).}
    \label{ElecOnesixteenth_Onetwelfth}
\end{figure}

The singlet pair-pair correlation function $D(r)$ for the $L_x = 192$ ladder is displayed in Fig.~\ref{PairOneeighth}. The numerical results are carefully extrapolated with truncation errors in DMRG calculations to remove the finite bond dimension effect.  We find $D(r)$ decays algebraically with oscillations in the period of CDW's wavelength $\lambda_c$, which makes the extraction of the exponent $K_{sc}$ hard. Apart from fitting the peaks using $D(r)\propto r^{-K_{sc 1}}$ as was typically done before \cite{PhysRevB.92.195139,lu2022ground}, we also fit the whole data by including an extra cosine term as $D(r) \propto \mathrm{A}_0 r^{-K_{sc 2}}+\mathrm{A}_1 \cos \left(2 \pi N_h/L_x+\phi\right) r^{-K_{sc 2}}$, which can be viewed as a first-order approximation to the oscillated pair-pair correlation function. We have checked both fits give the consistent results within error bar. We also checked the extracted value of $K_{sc}$ is converged with the size studied.

As can be seen from Fig.~\ref{PairOneeighth}(a), the pair-pair correlations clearly deviate from the algebraically decay behavior at the long distance near the boundary even after an extrapolation with truncation error. So to account for this effect, we exclude the very long-distance data for all the fits of pair-pair correlations. At the same time, we study long ladders which provide enough data for the fit process.

In the calculation of the pair-pair correlations, a more important factor is the choice of the position of the reference bond. Fig.~\ref{PairOneeighth}(b) shows the extracted $K_{sc}$ for different reference bonds (only correlations between vertical bonds are shown, the horizontal bonds have similar results). 
We find the boundary has a very large effect on $K_{sc}$. When moving the reference bond away from the boundary, $K_{sc}$ converges and the converged value satisfy the relationship $K_{sc} \cdot K_\rho = 1$ as predicted in the Luther Emery liquid theory.



%
\begin{table}[t]
    \caption{The dependence of the extracted parameters $K_\rho$ and $n_0$ (using Eq.~\eqref{charge_oscillation}) on the range of sites used ($\delta=1/4$, $L_x$ = 128). Also see a plot of $K_\rho$ versus the range of site used in Fig.~\ref{Fig2}(b).} 
    \includegraphics[width=0.3\textwidth]{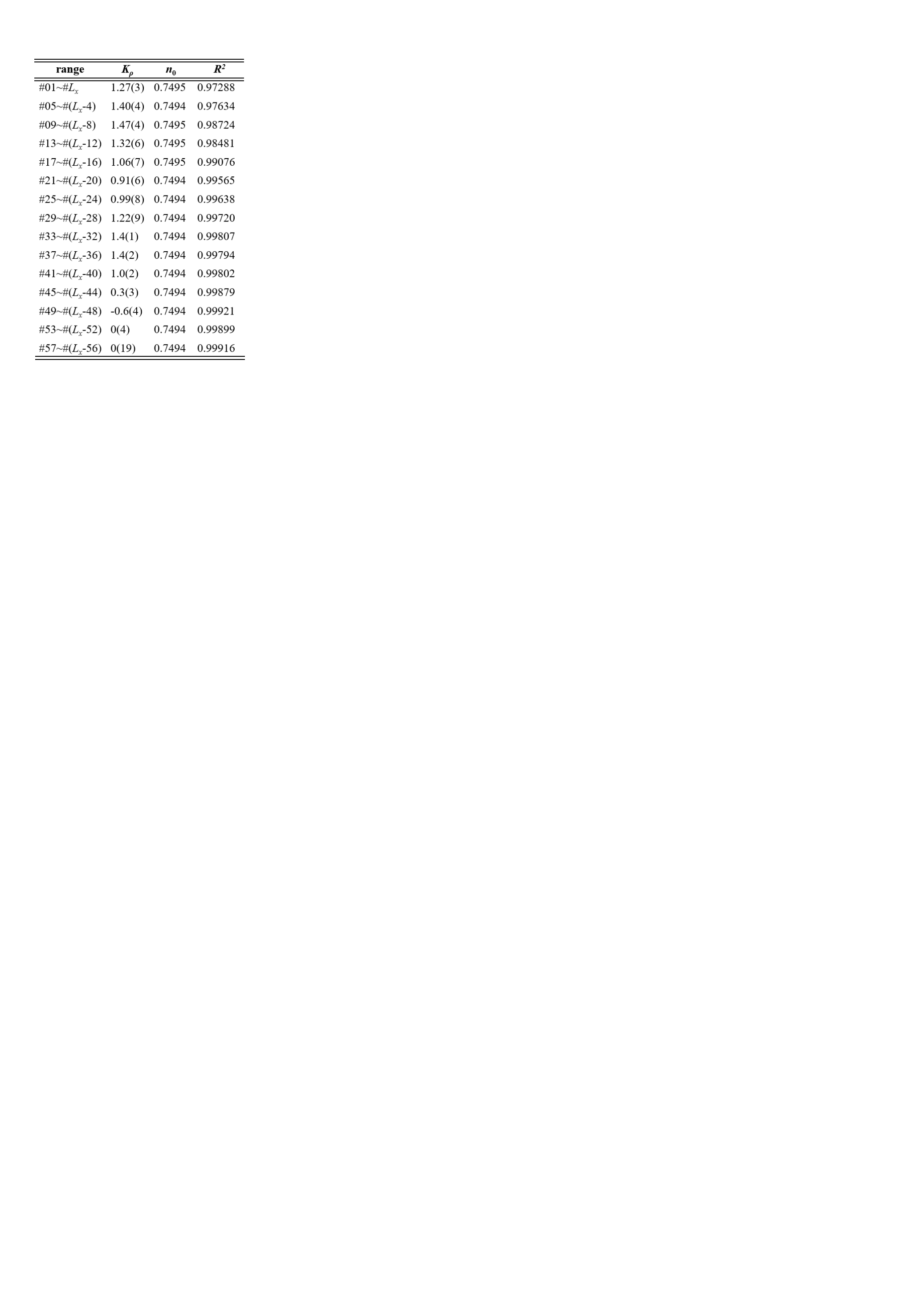}
    \label{ElecOscillationFit_onefourthL128}
\end{table}

\begin{figure*}[t]
    \includegraphics[width=0.3\textwidth]{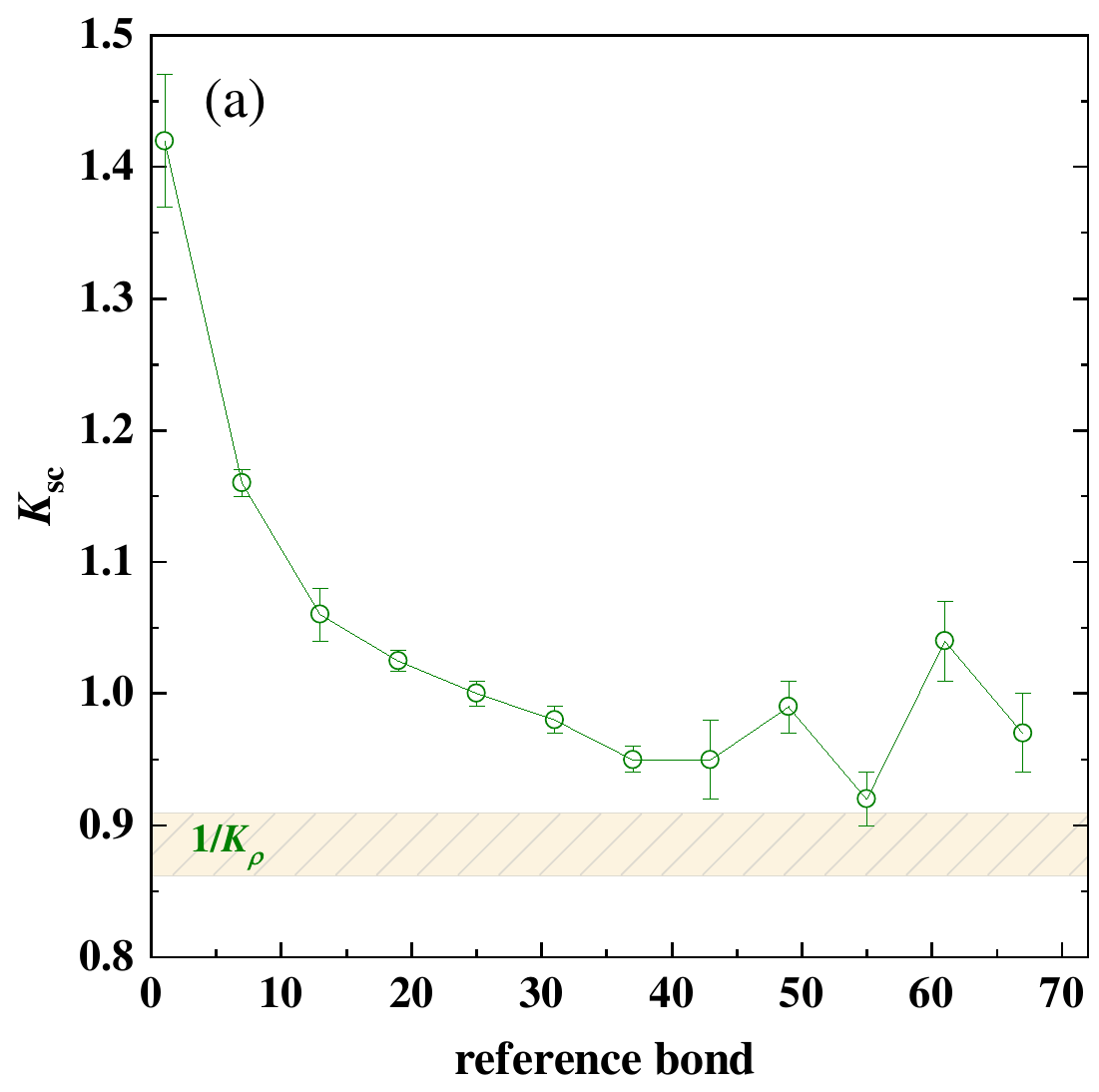}
    \includegraphics[width=0.3\textwidth]{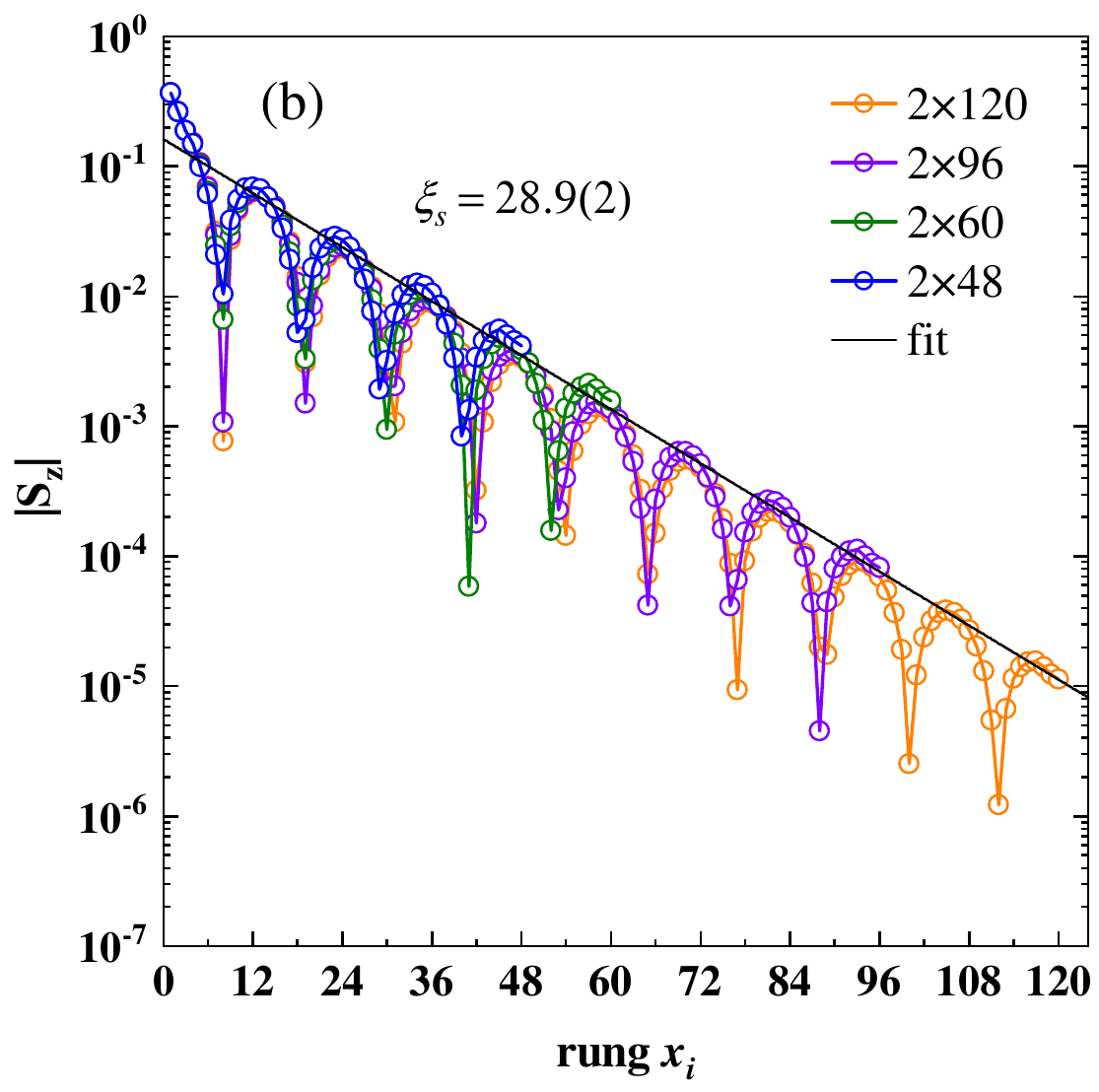}
    \includegraphics[width=0.3\textwidth]{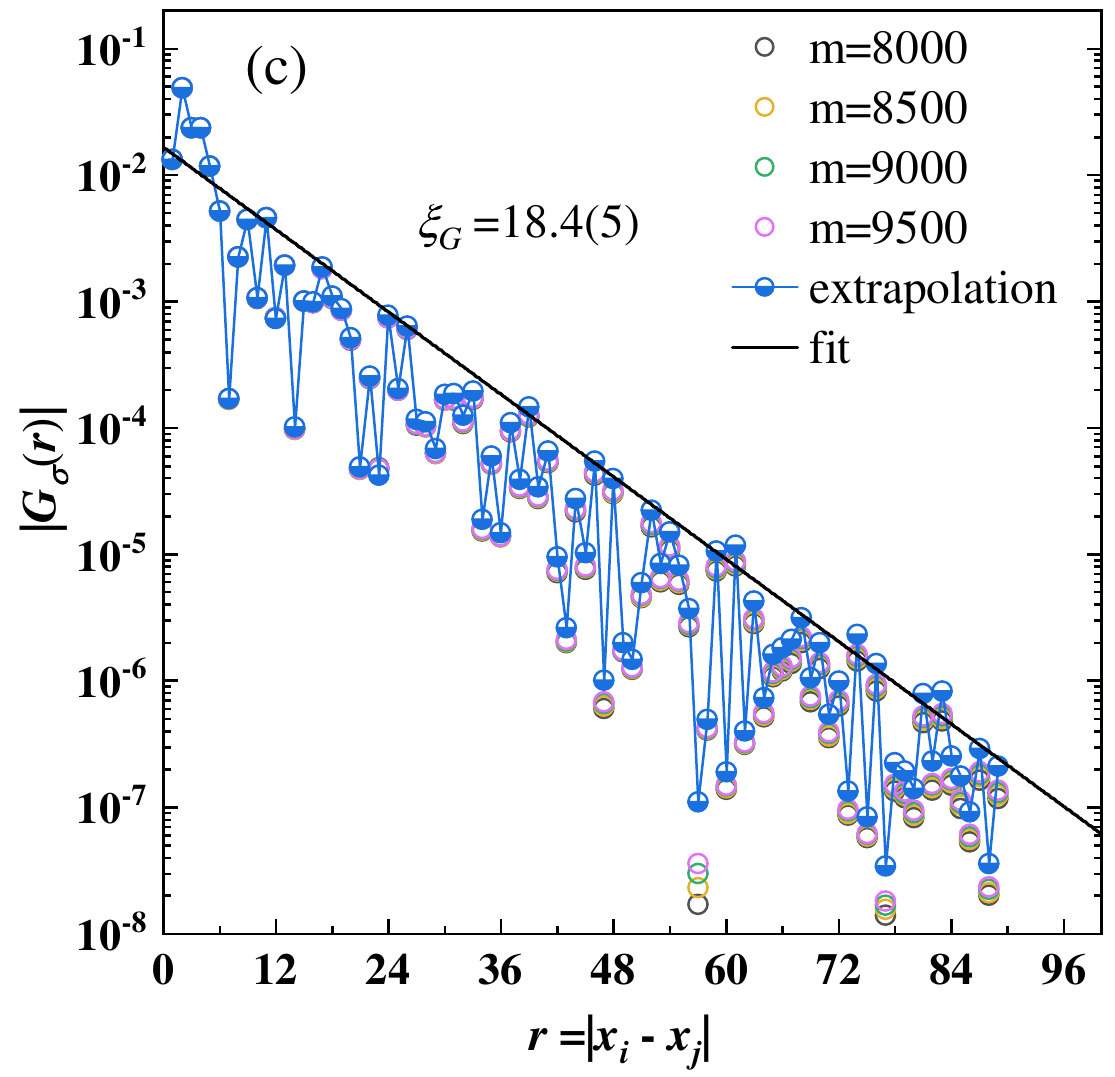}
    \includegraphics[width=0.3\textwidth]{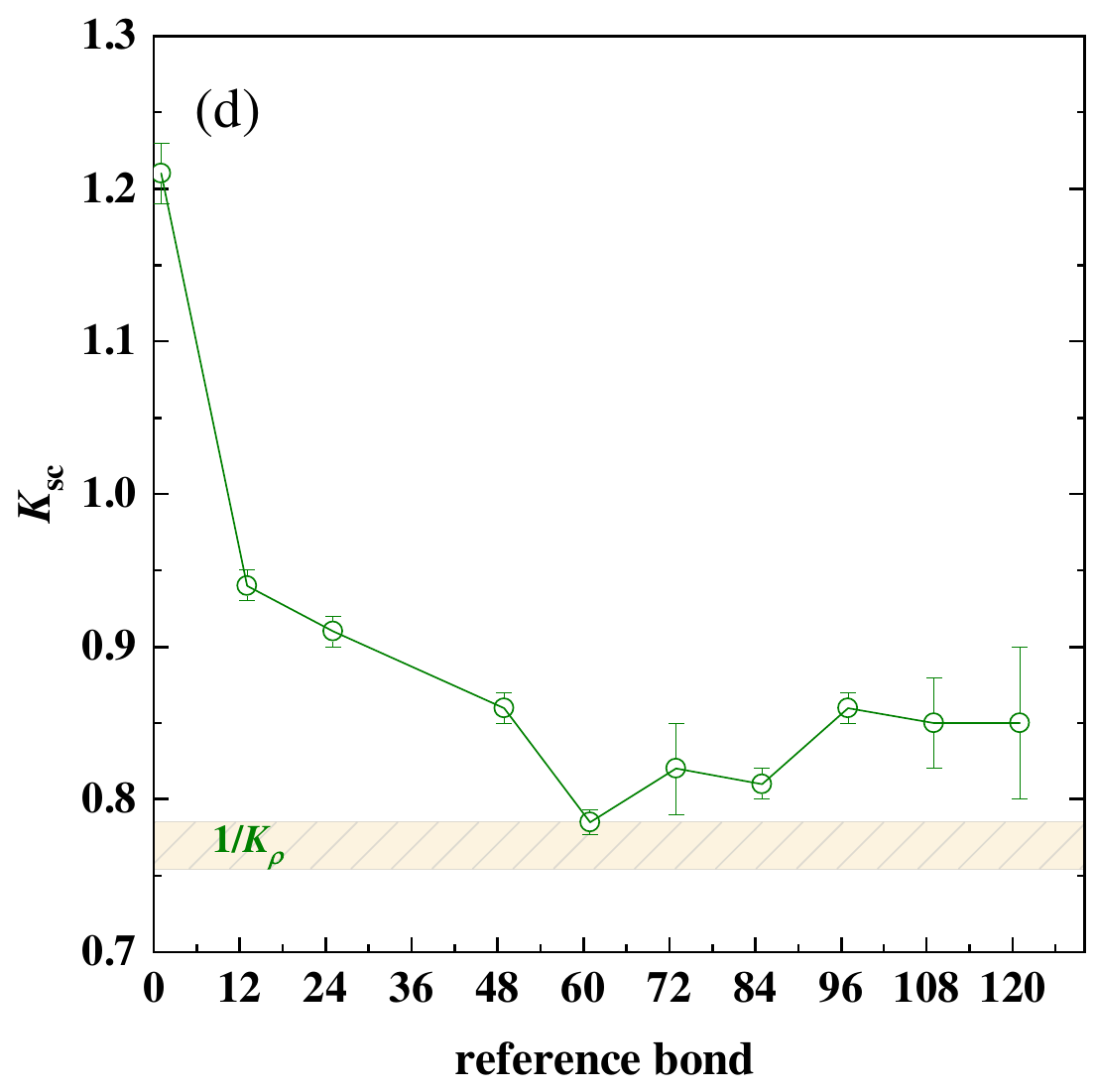}
    \includegraphics[width=0.3\textwidth]{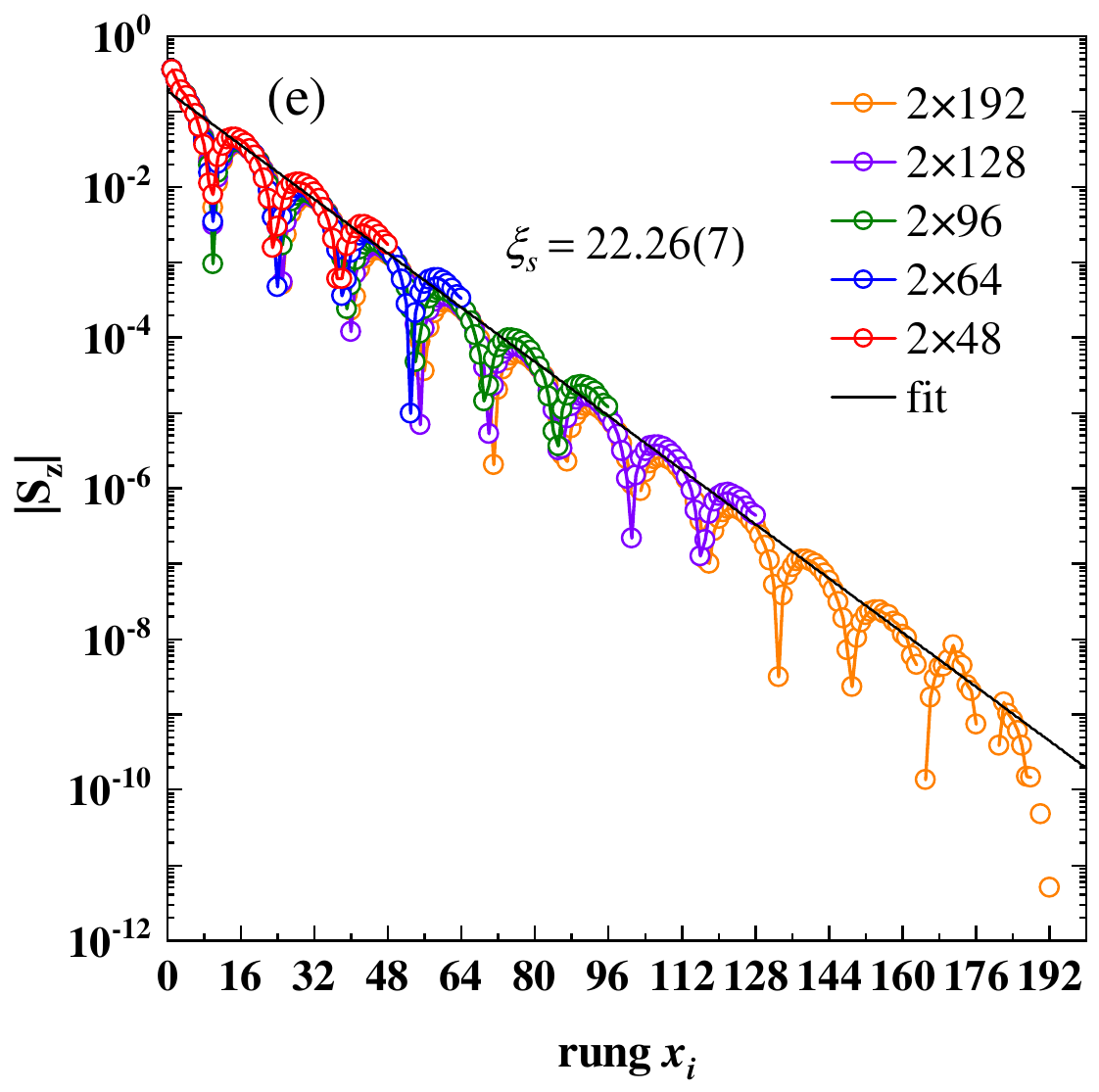}
    \includegraphics[width=0.3\textwidth]{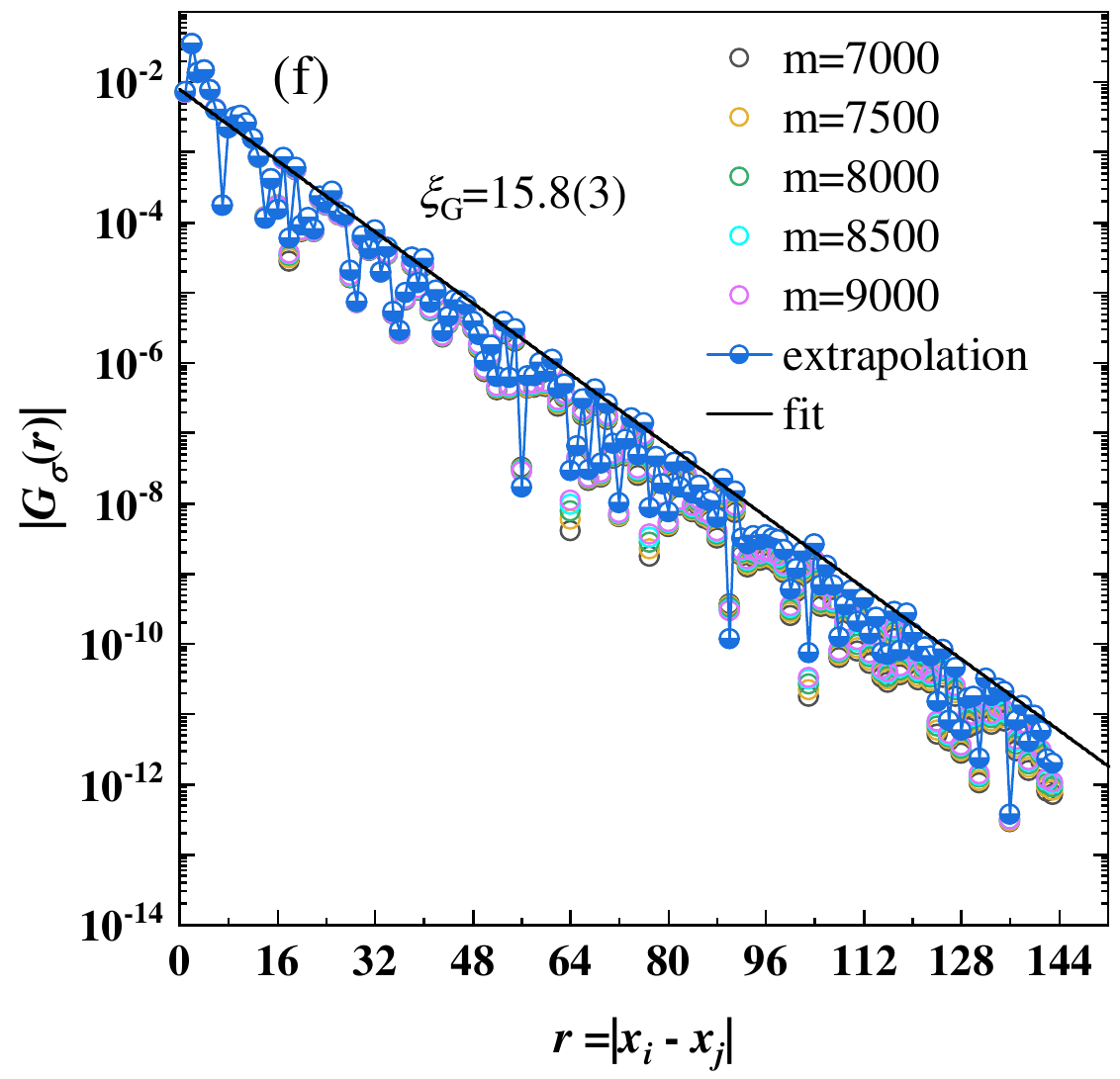}
    \caption{Results for $1/12$ (upper pane) and $1/16$ (lower pane) dopings. (a) and (d) show $K_{sc}$ as a function of the reference bond. The lengths of the systems are $L_x = 120$ for $\delta=1/12$ and $L_x = 192$ for $\delta=1/16$. The shaded area represents the value of $1/K_{\rho}$ with error bar. (b) and (e) show the absolute values of the local magnetization at $1/12$ and $1/16$ dopings with different lengths. Only the results from the extrapolation with truncation errors are shown. The solid lines denote exponential fits using $\left| \mathbf{\textit{S}}_\mathbf{\textit{z}}\right| \propto \mathrm{e}^{-x_i / \xi_{\mathrm{s}}}$ with $\xi_{\mathrm{s}}=28.9(2)$  for $\delta=1/12$($L_x=120$) and $\xi_{\mathrm{s}}=22.26(7)$ for $\delta=1/16$($L_x=192$). (c) and (f) show the single-particle Green's function for different numbers of kept states $m$ and the extrapolated result. The reference site is set at (31, 2) for $\delta=1/12$ ($L_x=120$) and (49, 2) for $\delta=1/16$ ($L_x=192$). Only peaked values are used in the fits for both the local magnetization and the single-particle Green's function.
    }
    \label{CorrelationOnetwelfth_Onesixteenth}
\end{figure*}

Fig.~\ref{OneeighthSz_Gf}(a) shows the absolute value of local magnetization for systems with different lengths. We find $\left|S_z\right|$ decays exponentially as $\left|S_z\right| \propto \mathrm{e}^{-x_i / \xi_{\mathrm{s}}}$ at long distances with a finite correlation length $\xi_{\mathrm{s}}=45.3(3)$, indicating the existence of a finite spin gap in the system which is consistent with the prediction of the Luther Emery liquid phase.
Moreover, the staggered spin density $(-1)^{x_i} \langle \hat{S}_{i}^z \rangle$ (not shown) has a spatial modulation with a wavelength twice that of the hole density and shows a $\pi$ phase flip at the hole concentrated position. The spin and hole modulations are consistent with the stripe phase \cite{tranquada1995evidence,doi:10.1073/pnas.96.16.8814}.

The single-particle Green's function is defined as $G_\sigma(r)=\langle \hat{c}_{(x_0,y),\sigma}^\dagger \hat{c}_{(x_0+r,y),\sigma} \rangle$ with $\sigma$ the spin index. We find $G_\sigma(r)$ is also sensitive to the position of the reference bond similar as the pair-pair correlations. We only show the converged results in this work. The results for $1/8$ doping is shown in Fig.~\ref{OneeighthSz_Gf}(b). $G_\sigma(r)$ decays exponentially as $\left| G_\sigma(r) \right| \propto \mathrm{e}^{-r / \xi_{\mathrm{G}}}$ with the correlation length $\xi_{\mathrm{G}} = 26.8(2)$.


From the above analysis of charge, pairing correlation, local magnetization, and single-particle Green's function, we conclude that the two-legged ladder at $1/8$ doping with $U = 8$ belongs to the Luther-Emery liquid phase and the exponents of charge and superconducting correlations satisfy the relationship $K_{sc} \cdot K_\rho =1$.

\subsection{$1/12$ and $1/16$ dopings}


We perform similar calculations and analyses for $1/12$ and $1/16$ dopings in this subsection.
The charge density profiles for 1/12 and 1/16 dopings are shown in Figs.~\ref{ElecOnesixteenth_Onetwelfth}(a) and (b), which also display the Friedel oscillations with $\lambda_c = 1/\delta$. We find the amplitude of charge density oscillations is reduced with the decrease of doping level. The exponents increases from $K_\rho = 1.13(3)$ to $K_\rho = 1.28(3)$ from $\delta = 1/12$ to $\delta = 1/16$ (see the insets of Fig.~\ref{ElecOnesixteenth_Onetwelfth} and Table~\ref{Summary}). This behavior is consistent of the prediction that $K_\rho \rightarrow 2$ with $\delta \rightarrow 0$ \cite{PhysRevB.59.R2471,PhysRevB.63.195106}.

Following the fitting procedures in the 1/8 doping case, we also extract $K_{sc}$ form the pair-pair correlations by varying the position of reference bond as shown in Figs.~\ref{CorrelationOnetwelfth_Onesixteenth}(a) and \ref{CorrelationOnetwelfth_Onesixteenth}(d).
The converged values for $K_{sc}$ are $0.95(3)$ and $0.85(5)$ for $1/12$ and $1/16$ dopings respectively. For these dopings,
$K_{sc} < K_\rho$ indicating superconducting correlation is dominant at low dopings \cite{PhysRevB.92.195139}. Similar as the $1/8$ doping case, the relationship $K_{sc} \cdot K_\rho = 1$ is also likely to be satisfied, which is consistent with the prediction from Luther-Emery liquid theory.

The absolute value of local magnetization and single-particle Green's function for 1/12 and 1/16 dopings are shown in Figs.~\ref{CorrelationOnetwelfth_Onesixteenth}(b), ~\ref{CorrelationOnetwelfth_Onesixteenth}(c) and Figs.~\ref{CorrelationOnetwelfth_Onesixteenth}(e), \ref{CorrelationOnetwelfth_Onesixteenth}(f).
Both of them decay exponentially with distance and the correlation lengths are smaller than $1/8$ cases, indicating the spin and single-particle gap increase with the decrease of doping.



%
\begin{table}[t]
    \caption{The dependence of the extracted parameters $K_\rho$ and $n_0$ (using Eq.~\eqref{charge_oscillation} on the range of sites used ($\delta=1/3$, $L_x$ = 96)). Also see a plot of $K_\rho$ versus the range of site used in Fig.~\ref{Fig2}(c).}
    \includegraphics[width=0.3\textwidth]{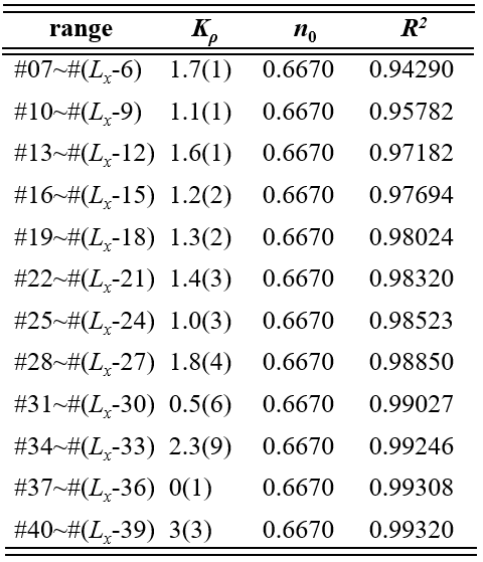}
    \label{ElecOscillationFit_onethirdL96}
\end{table}

\begin{figure}[b]
    \includegraphics[width=0.4\textwidth]{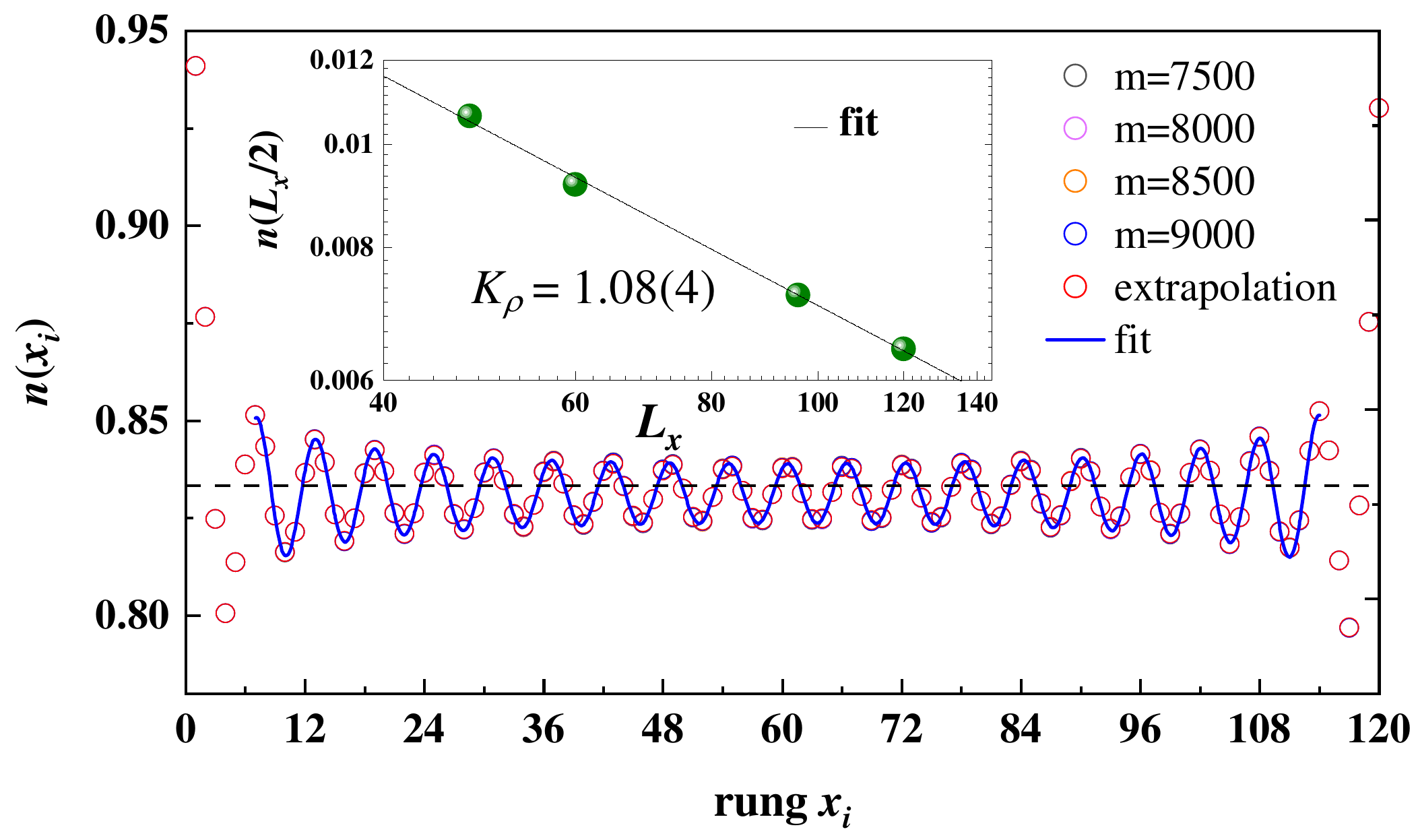}
    \caption{Density profile at $1/6$ doping. The local rung density $n(x)$ for different kept states $m$ and the extrapolated with truncated errors results are shown. The length of the system is $L_x =120$ at $1/6$ doping. The solid line is the fitting curve using Eq.~\eqref{charge_oscillation} to extract $K_\rho$. The dashed horizontal line represents the average electron density. The inset is the finite-size scaling of $\delta n(L_x/2)$ as a function of the system size $L_x$ using Eq.~(\ref{finite_scale}).}
    \label{ElecOnesixthL120}
\end{figure}

\subsection{$1/6$ doping level}

\begin{figure}[t]
    \includegraphics[width=0.4\textwidth]{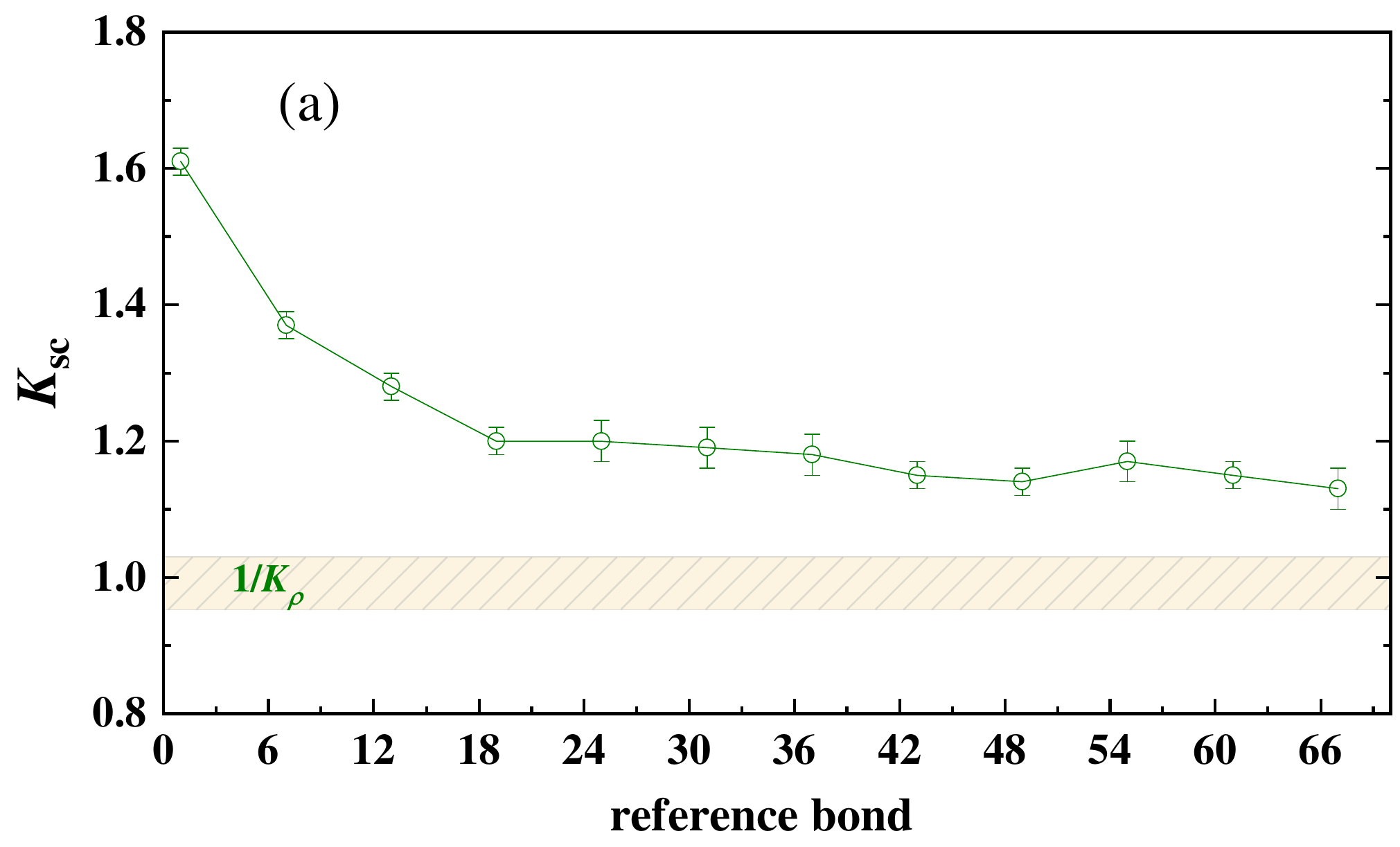}
    \includegraphics[width=0.4\textwidth]{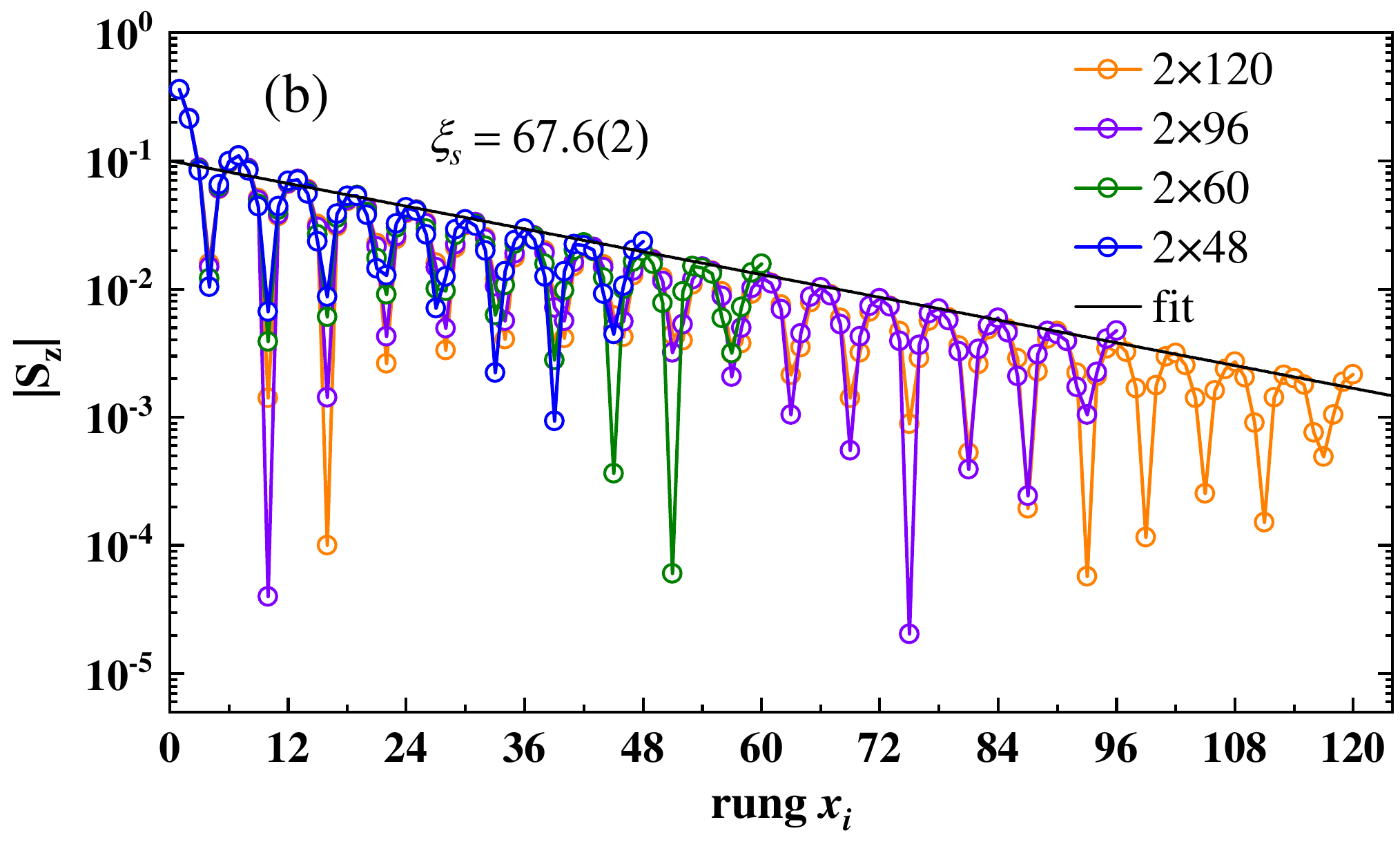}
    \includegraphics[width=0.4\textwidth]{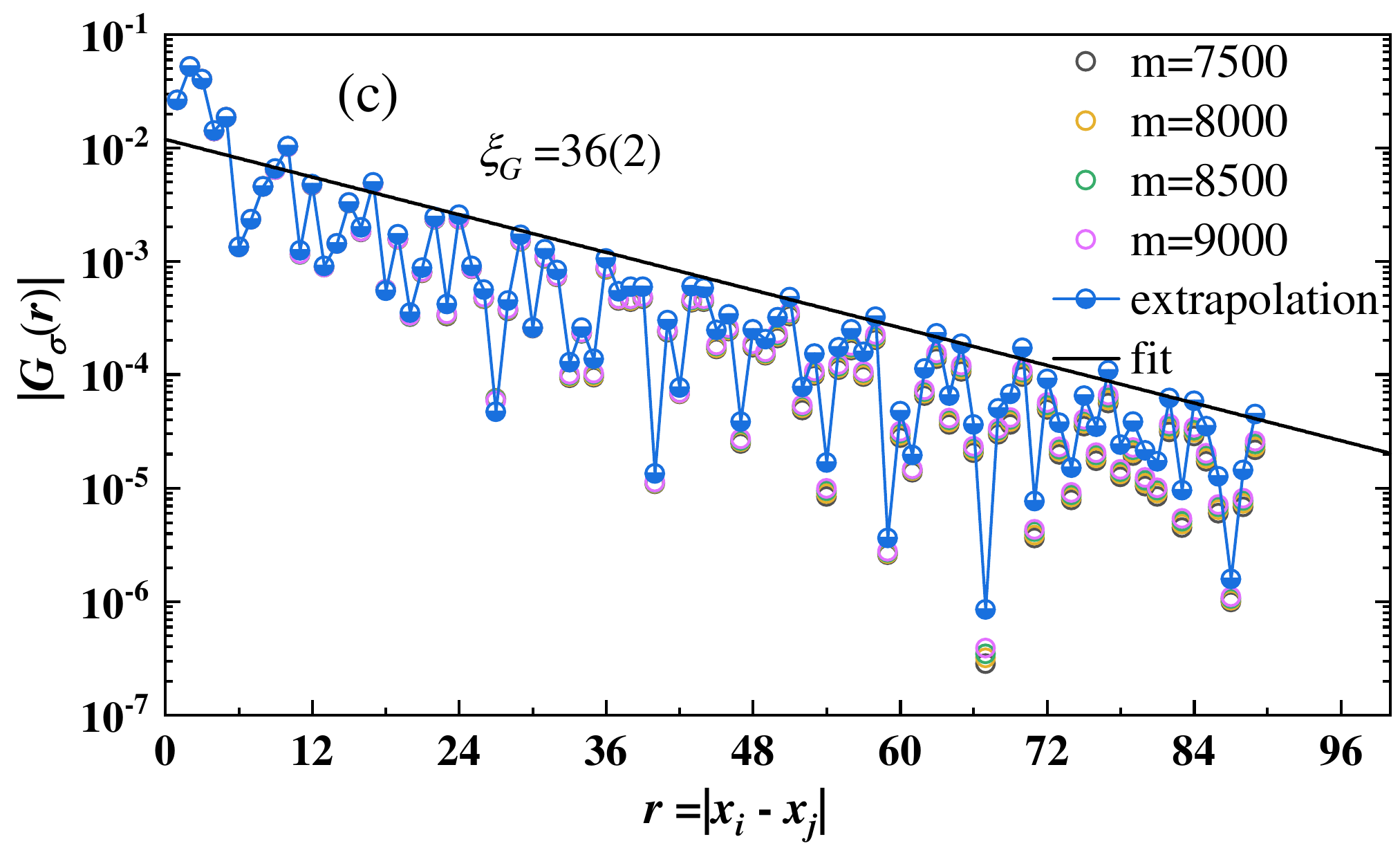}
    \caption{Results for $1/6$ doping level. (a) shows $K_{sc}$ as a function of the reference bond. The length of the system is $L_x = 120$. The shaded area represents the value of $1/K_{\rho}$ with error bar. (b) shows the absolute values of the local magnetization at 1/6 doping for systems with different lengths. Only the results from the extrapolation with truncation errors are shown. The solid lines denote exponential fitting using $\left| \mathbf{\textit{S}}_\mathbf{\textit{z}}\right| \propto \mathrm{e}^{-x_i / \xi_{\mathrm{s}}}$ with $\xi_{\mathrm{s}}=67.6(2)$  for $L_x=120$. (c) shows the single-particle Green's function for different numbers of kept states $m$ and the extrapolated result. The reference site is set at (31, 2). Only peaked values are used in the exponential fits for both the local magnetization and the single-particle Green's function.
    }
    \label{Onesixth}
\end{figure}

In this subsection, we move to the $1/6$ doping case.
Fig.~\ref{ElecOnesixthL120} shows the charge density profiles for a $L_x = 120$ ladder, which is also characterized by an algebraic quasi-long range order with period $\lambda_c=1/\delta=6$. As we can see in the inset of Fig.~\ref{ElecOnesixthL120}, $K_\rho$ extracted from two approaches are consistent with $K_\rho \approx 1.01(4)$. Following the procedure described above, we show the variance of $K_{sc}$ with the position of reference bond in Fig.~\ref{Onesixth}(a). We find $K_\rho < K_{sc}$ (also see Table~\ref{Summary}), suggesting the charge correlation is dominating at 1/6 doping. We find that charge and superconducting exponents are likely to violate the relationship $K_{sc} \cdot K_\rho = 1$.
The local magnetization and single-particle Green's function are shown in Figs.~\ref{Onesixth}(b) and ~\ref{Onesixth}(c). Both of them decay exponentially with correlation lengths $\xi_{\mathrm{s}}=67.6(2)$ and $\xi_G=36(2)$.



\subsection{$1/3$ and $1/4$ dopings}

\begin{figure}[h]
    \includegraphics[width=0.4\textwidth]{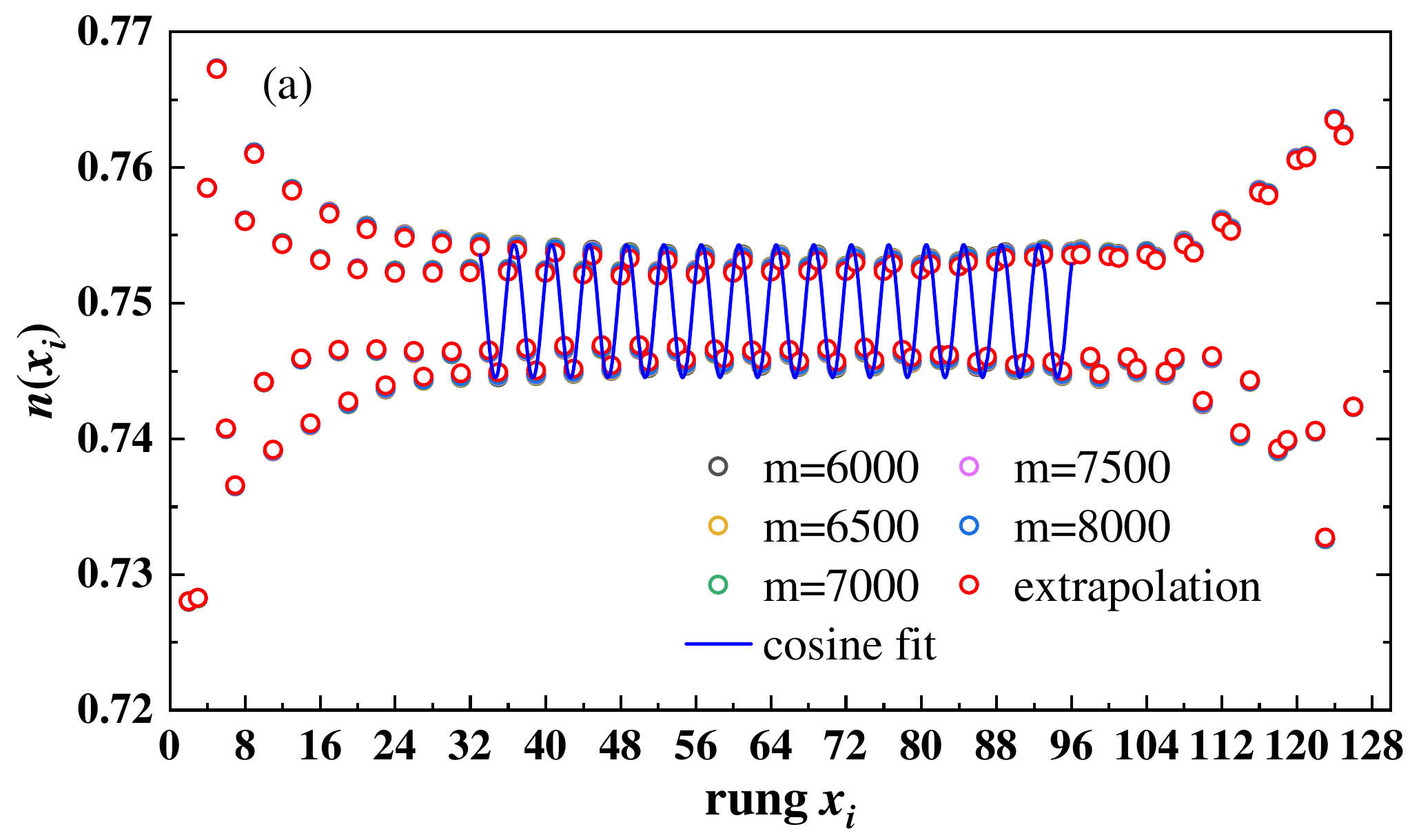}
    \includegraphics[width=0.4\textwidth]{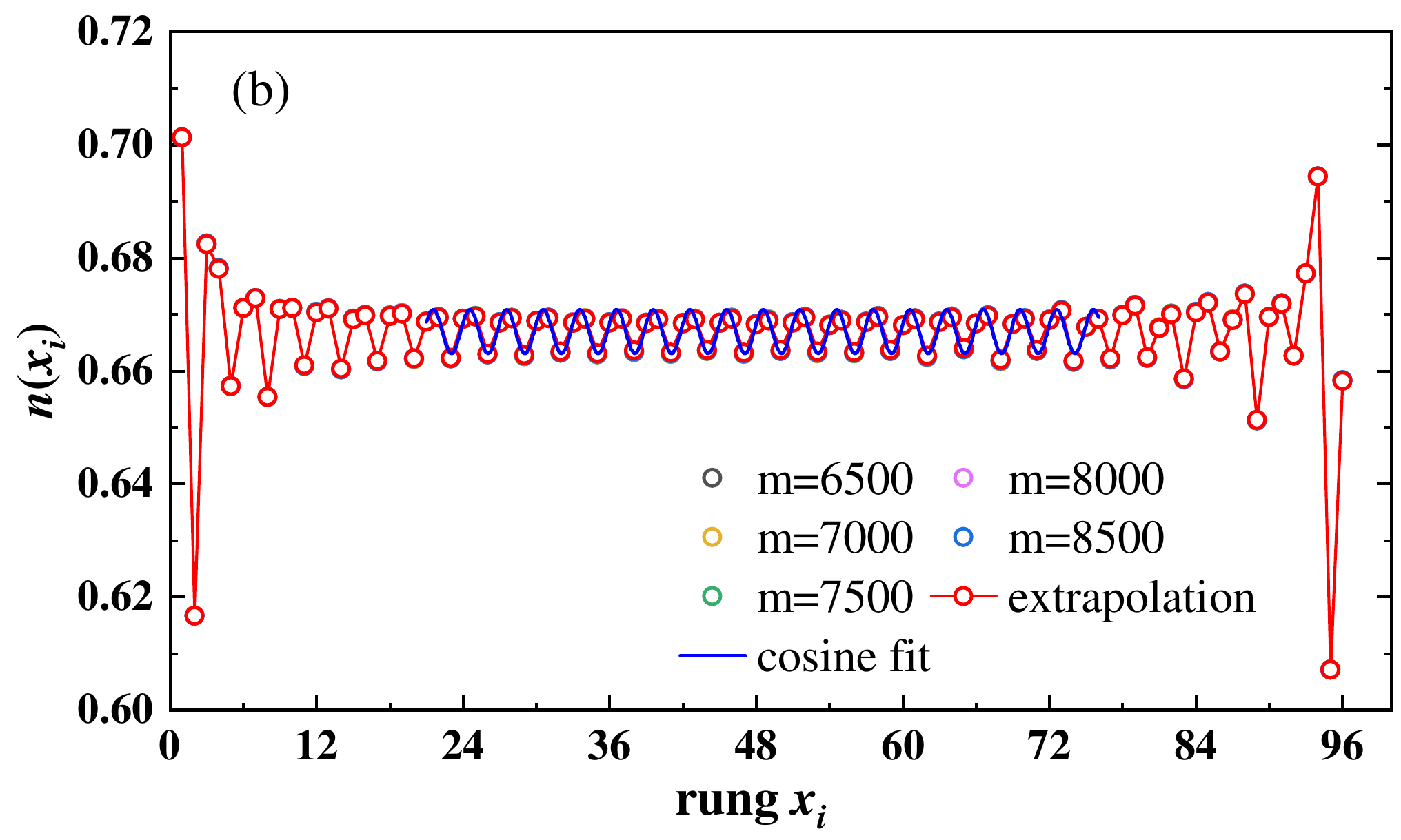}
    \caption{Density profiles at (a) $1/4$ and (b) $1/3$ dopings. The local rung density $n(x)$ for different kept states $m$ and the extrapolated with truncated errors results are shown. The lengths of the systems are $L_x =128$ for $1/4$ doping and $L_x =96$ for $1/3$ doping, respectively. The solid lines are the fitting curves using a cosine function. }
    \label{ElecOnefourth_Onethird}
\end{figure}


Fig.~\ref{ElecOnefourth_Onethird}(a) shows the charge density profile at $1/4$ doping for a $L_x=128$ ladder. We can see that the charge modulation has a persistent amplitude with period $\lambda_c = 1/\delta$ which doesn't decay in the bulk. The extracted $K_\rho$ using Eq.~\eqref{charge_oscillation} are shown in Fig.~\ref{Fig2}(b) and Table~\ref{ElecOscillationFit_onefourthL128}. The value of $K_\rho$ varies dramatically with the data range used in the fit. The fact that $K_\rho$ has very large error bars and oscillates across zero indicates the failure of the fit using Eq.~\eqref{charge_oscillation}, which suggests the charge order is likely long-ranged in the system.

The pair-pair correlation for $\delta=1/4$ are shown in Figs.~\ref{Onefourth}(a) and ~\ref{Onefourth}(b). Following the same procedure above, we find $K_{sc} = 1.05(6)$ at $\delta = 1/4$.
We show the local magnetization in both semi-logarithmic and double-logarithmic scale in Fig.~\ref{Onefourth}(c) and (d). We find the data is better described by a exponential decay fit with a very large correlation length $\xi_{\mathrm{s}} =107(2)$ instead of a power-law decay. The single-particle Green's function is found to decay exponentially with a correlation length  $\xi_{\mathrm{G}} =50(2)$ (see Fig.~\ref{Onefourth}(e) and (f)).

The results for $1/3$ doping are similar as $1/4$ doping. The charge order is likely to be a long-ranged one as shown in Fig.~\ref{ElecOnefourth_Onethird}(b). In Fig.~\ref{Fig2}(c) and Table~\ref{ElecOscillationFit_onethirdL96} , we find the extracted value of $K_\rho$ also varies dramatically with the data range. The extracted $K_{sc} = 1.10 (8)$ as shown in Figs.~\ref{Onethird}(a) and (b). The spatial decay of local magnetization can be well fitted using either an exponential law or a power law but with correlation length equal to the size of the system (see Figs.~\ref{Onethird}(c) and ~\ref{Onethird}(d), indicating the spin gap is likely to be closed at $1/3$ doping. The decay of single-particle Green's function is found to be power-law like, indicating the closing of the gap at $1/3$ doping (see Fig.~\ref{Onethird}(e) and (f)).


\section{Conclusions}
\label{sec_con}
We revisit the ground state of the Hubbard model on 2-legged ladders with DMRG. We find $K_{sc}$ extracted from the algebraic decay of pair-pair correlation depends strongly on the position of reference bond. We obtain the most accurate exponent $K_\rho$ and $K_{sc}$ with DMRG after systematically treating the effects from boundary conditions, finite sizes, and truncation errors in DMRG. We confirm that the 2-legged Hubbard model is in the Luther-Emery liquid phase with $K_\rho \cdot K_{sc} = 1$ from tiny doping near half-filling to 1/8 hole doping, resolving the long-time controversy. When the doping is increases to $\delta \gtrapprox 1/6$, the behaviors of charge, pairing, and spin correlations don't change qualitatively, but the relationship $K_{sc} \cdot K_\rho = 1$ is likely to be violated. With the further increase of the doping to $\delta = 1/3$, the quasi long-ranged charge correlation turns to a long-ranged charge order and the spin gap is closed, while the pair-pair correlation still decays algebraically.
In \cite{shen2022comparative}, the Hubbard ladder at $1/3$ doping with next-nearest-neighbouring hopping $t'$ and a larger $U= 12$ was studied \cite{shen2022comparative}. The charge correlation was found to decay algebraically, while the spin and single-particle excitations are gapped \cite{shen2022comparative}. The comparison of the results in \cite{shen2022comparative} and in this work indicates the boundary of different phases depend on the details of the parameters in the model. Our work provides a standard way to analyze the correlation functions when studying systems with open boundaries.

\begin{figure*}[h]
    \includegraphics[width=0.4\textwidth]{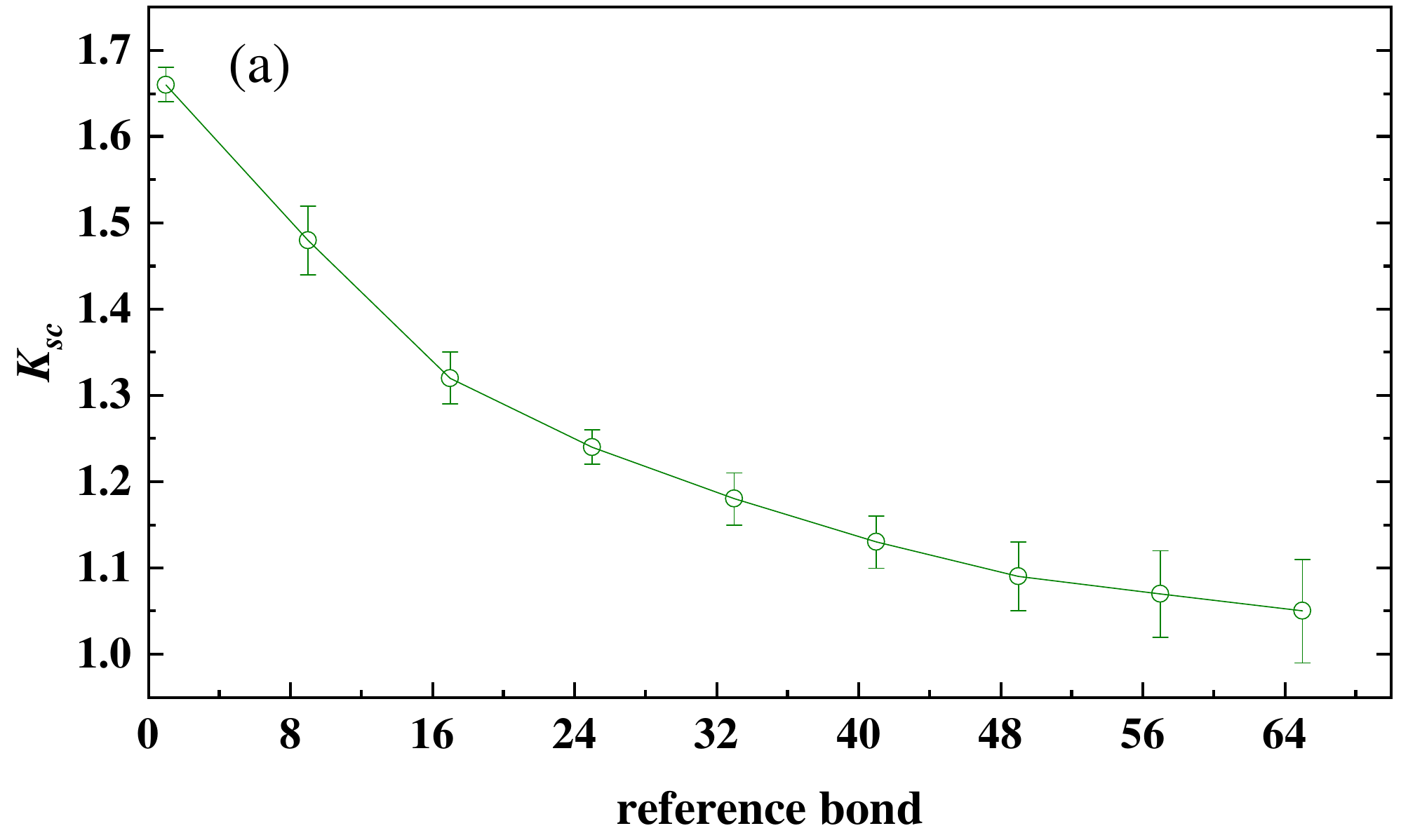}
    \includegraphics[width=0.4\textwidth]{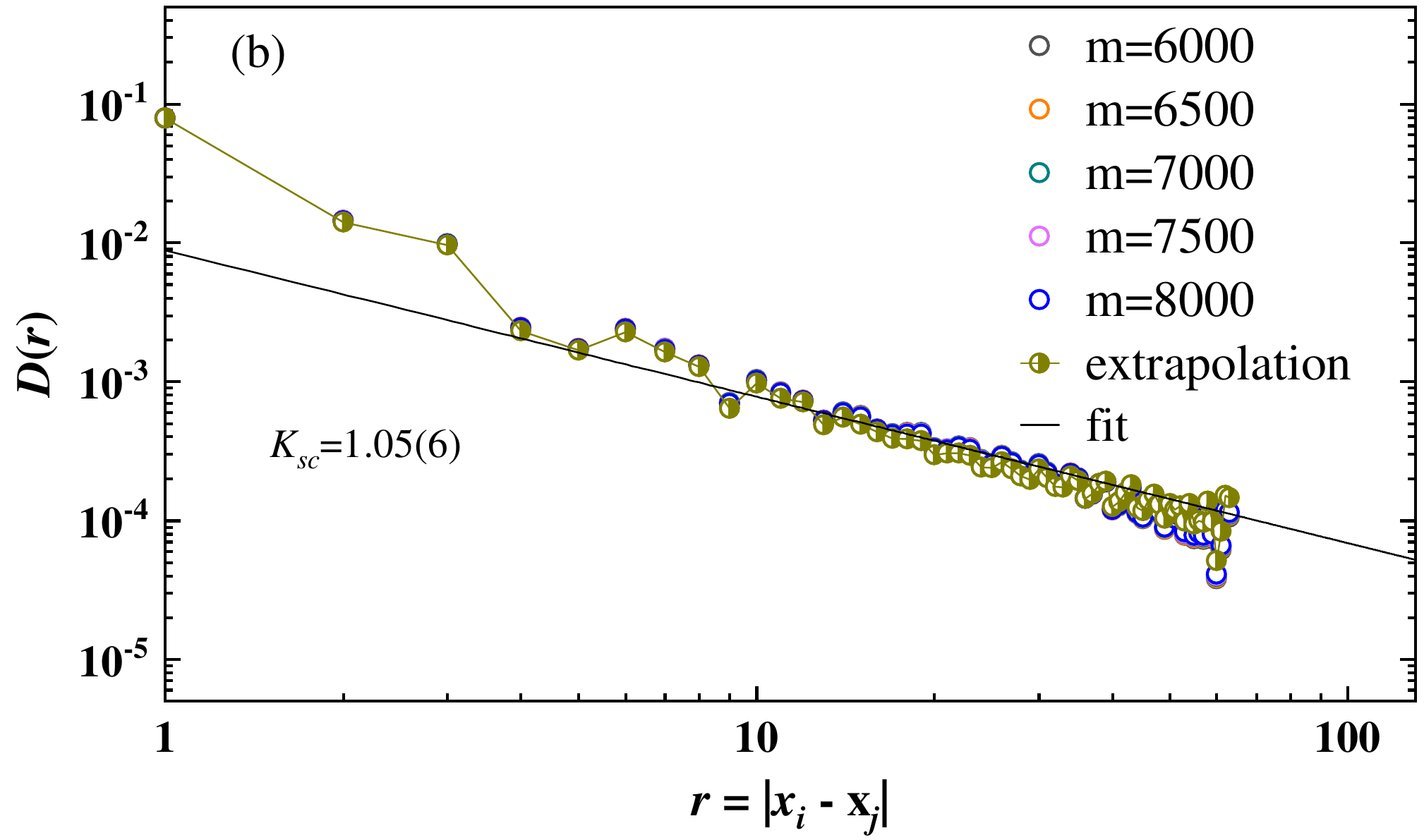}
    \includegraphics[width=0.4\textwidth]{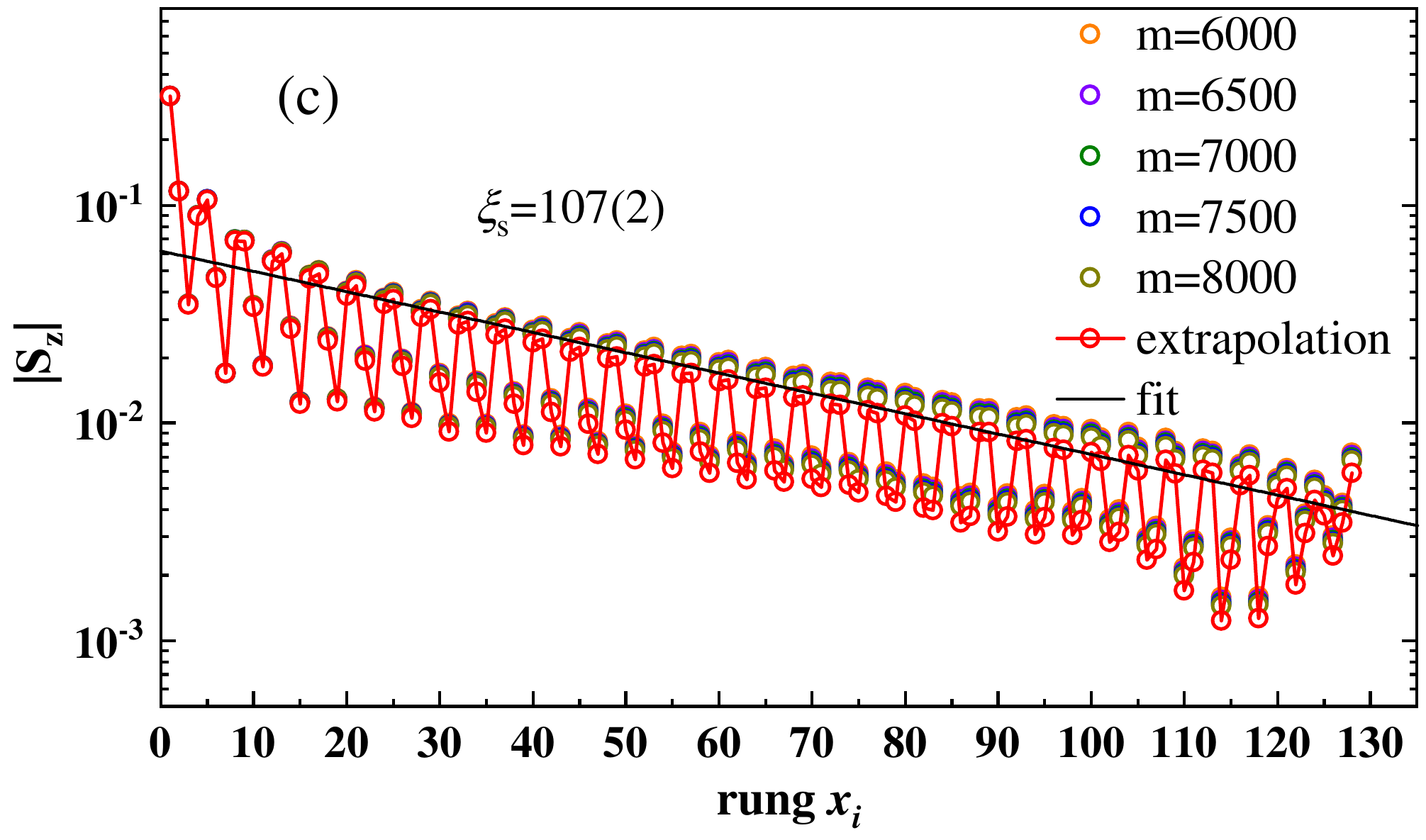}
    \includegraphics[width=0.4\textwidth]{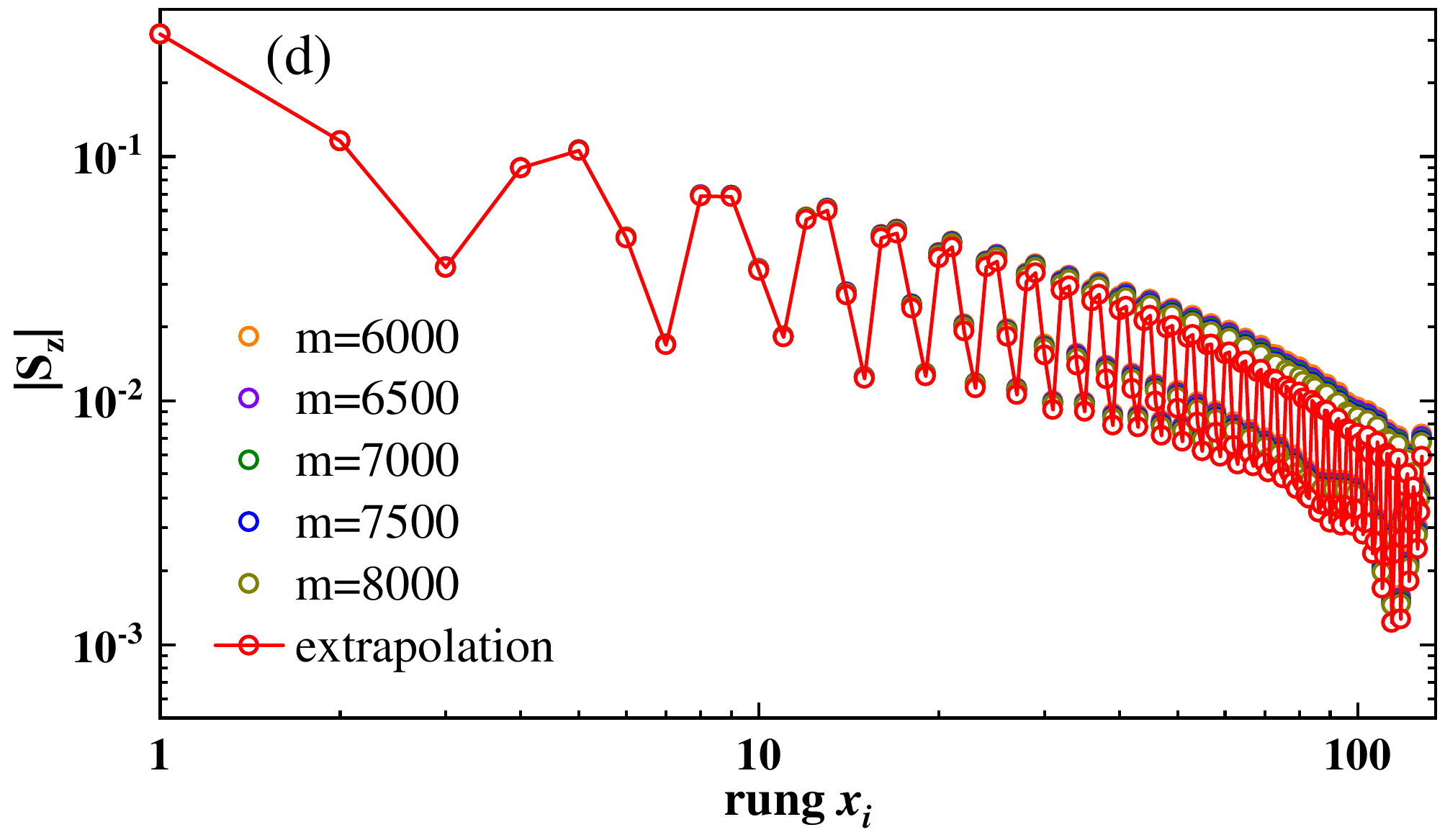}
    \includegraphics[width=0.4\textwidth]{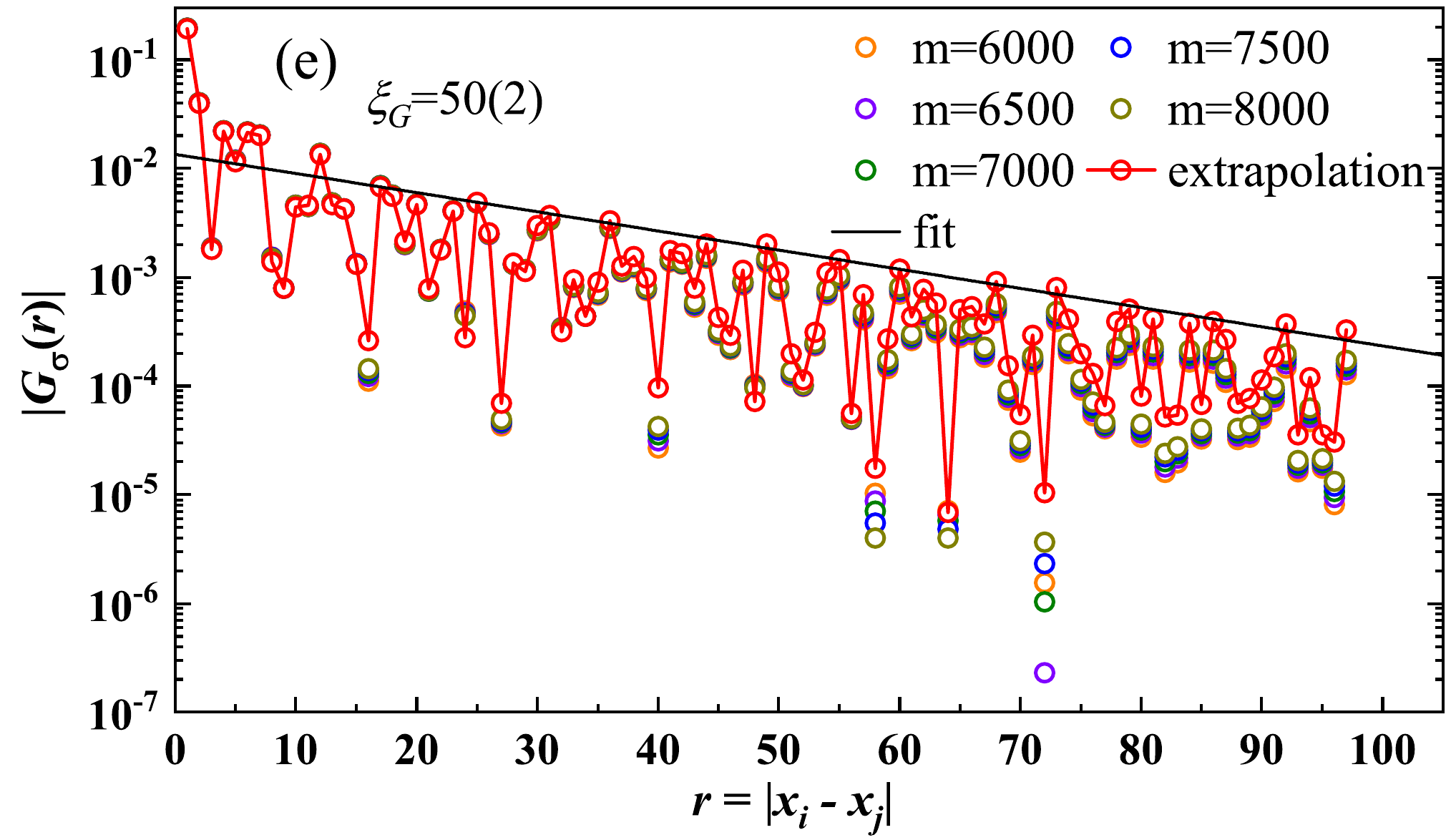}
    \includegraphics[width=0.4\textwidth]{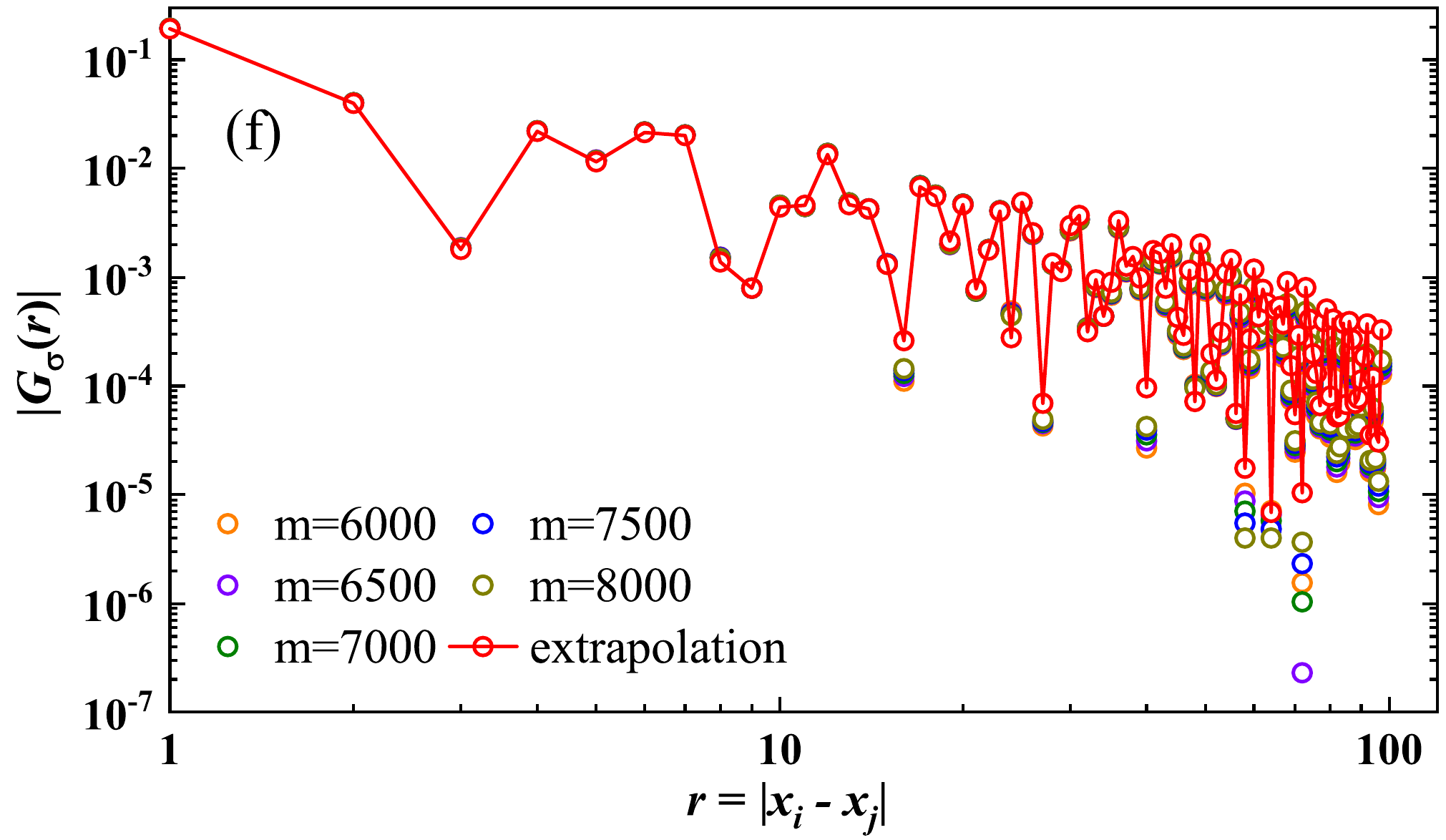}
    \caption{Pair-pair correlation, local magnetization and single-particle Green's function at $1/4$ doping. (a) $K_{sc}$ as a function of the reference bond. The length of the system is $L_x = 128$. (b) A typical fit of pair-pair correlation function. The reference bond is set at the 65th vertical bond. Absolute values of local magnetization (c, d) and single-particle Green's function (e, f). The results from finite kept state and the extrapolated with truncation errors results are shown.  The plots in (c, e) and (d, f) are shown in the semi- and double-logarithmic axes, respectively. The solid line in (c) denotes exponential fitting using $\left| \mathbf{\textit{S}}_\mathbf{\textit{z}}\right| \propto \mathrm{e}^{-x_i / \xi_{\mathrm{s}}}$ with $\xi_{\mathrm{s}}=107(2)$. By comparing (c) and (d), we find the decay of the local magnetization is likely to be exponential, even though the extracted correlation length is comparable to the length of the system. The single-particle Green's function is also more likely to exhibit exponential decay. Only peaked values are used in the fits.}
    \label{Onefourth}
\end{figure*}

\begin{figure*}[h]
    \includegraphics[width=0.4\textwidth]{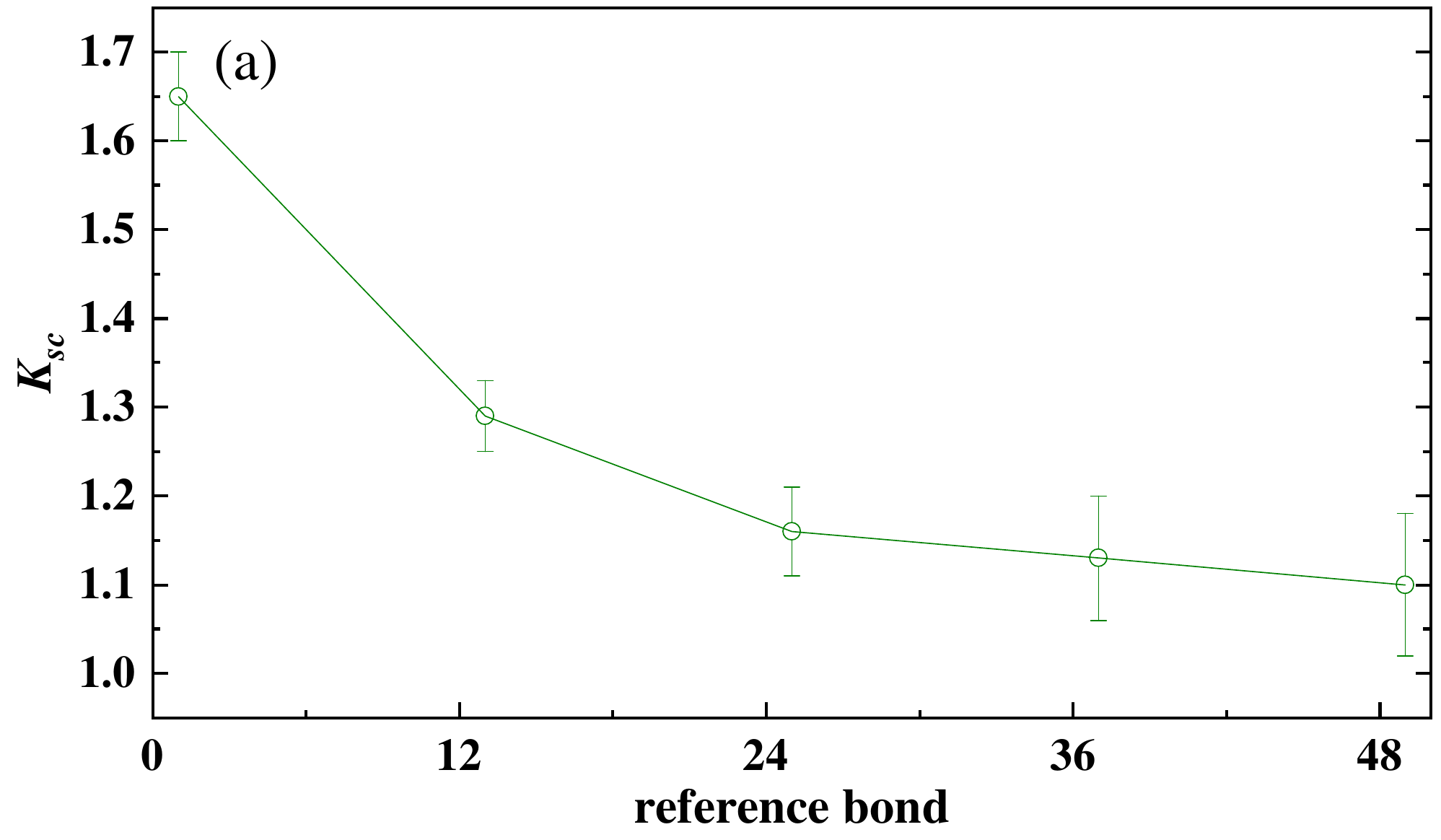}
    \includegraphics[width=0.4\textwidth]{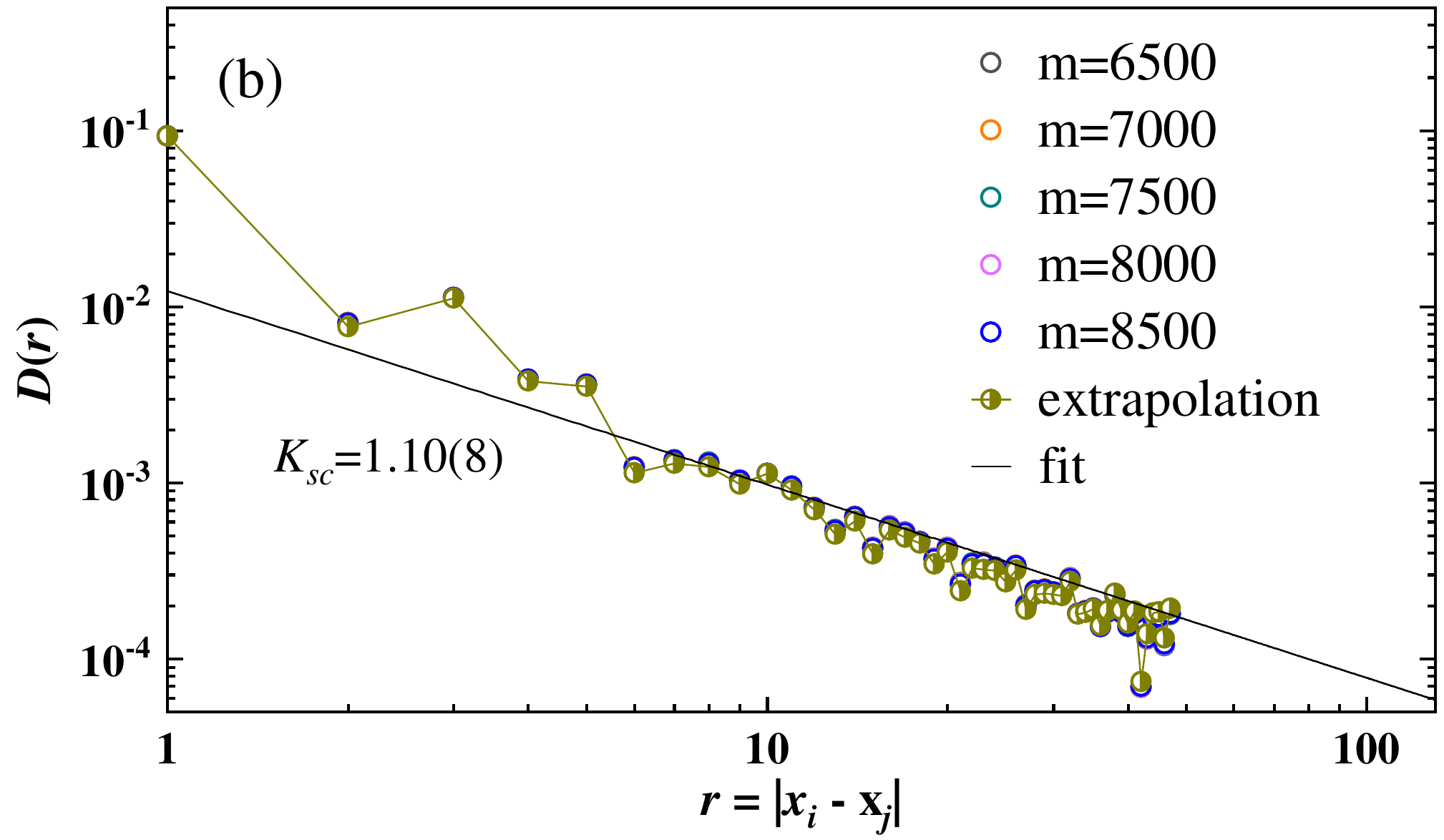}
    \includegraphics[width=0.4\textwidth]{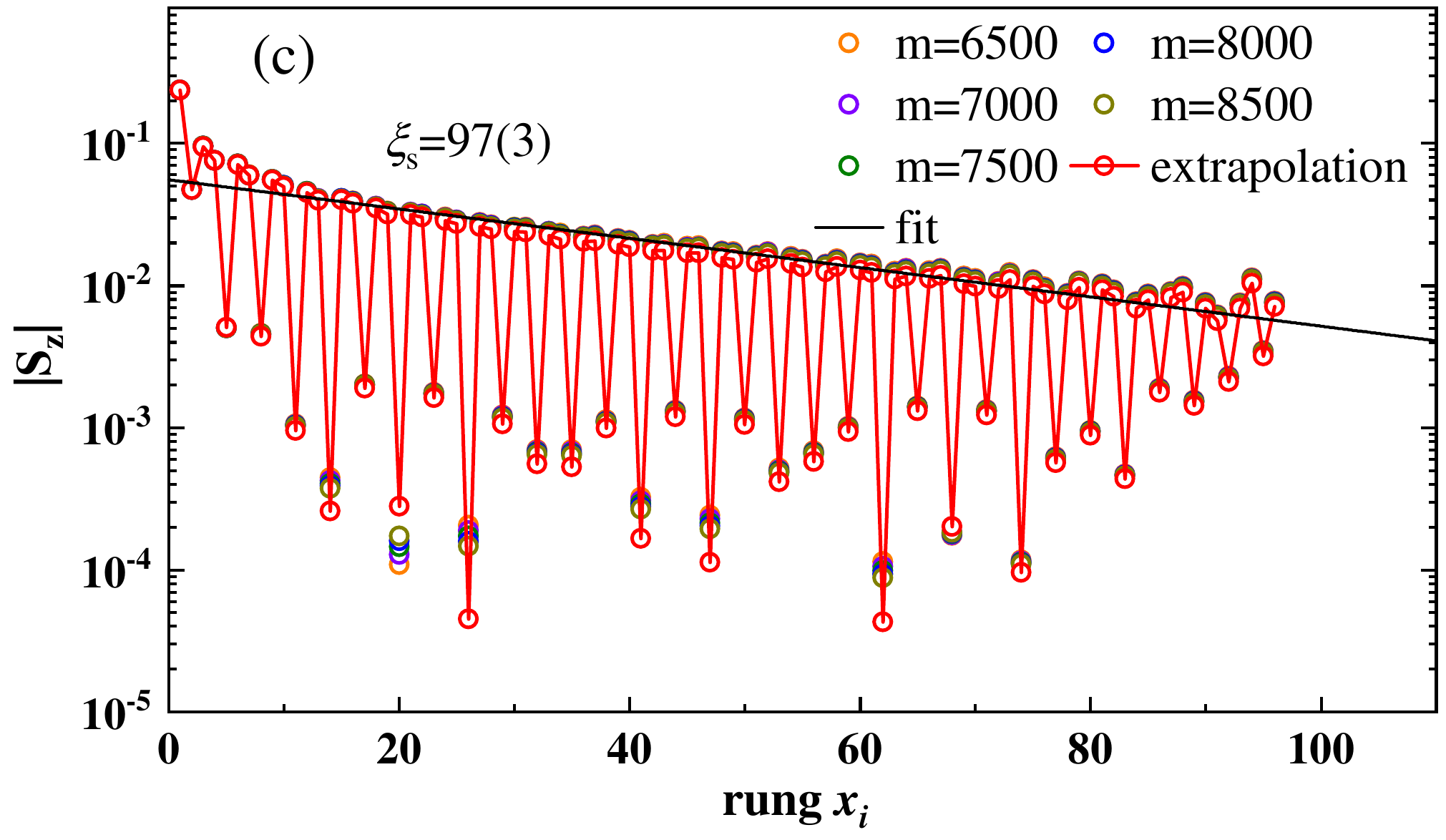}
    \includegraphics[width=0.4\textwidth]{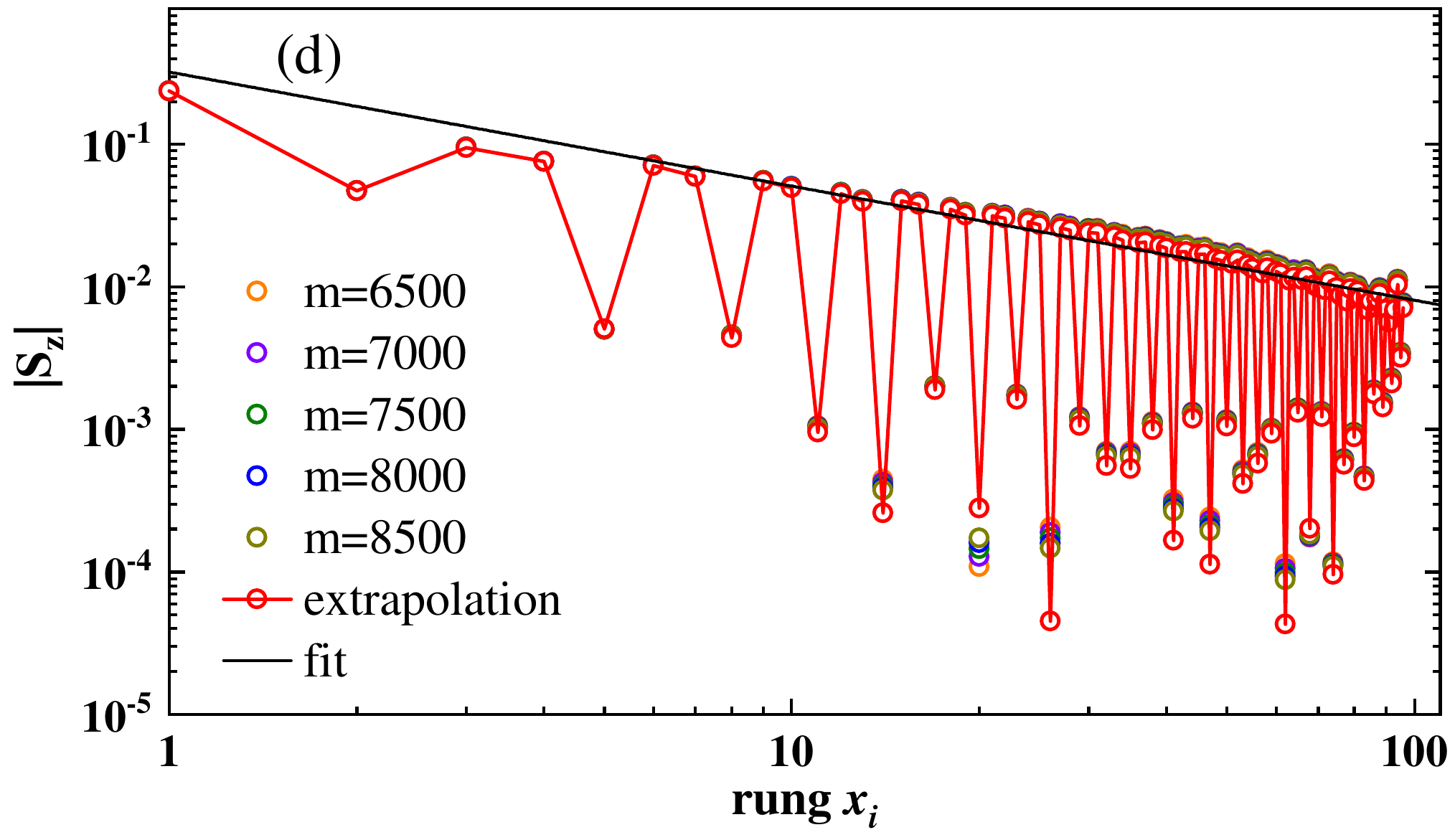}
    \includegraphics[width=0.4\textwidth]{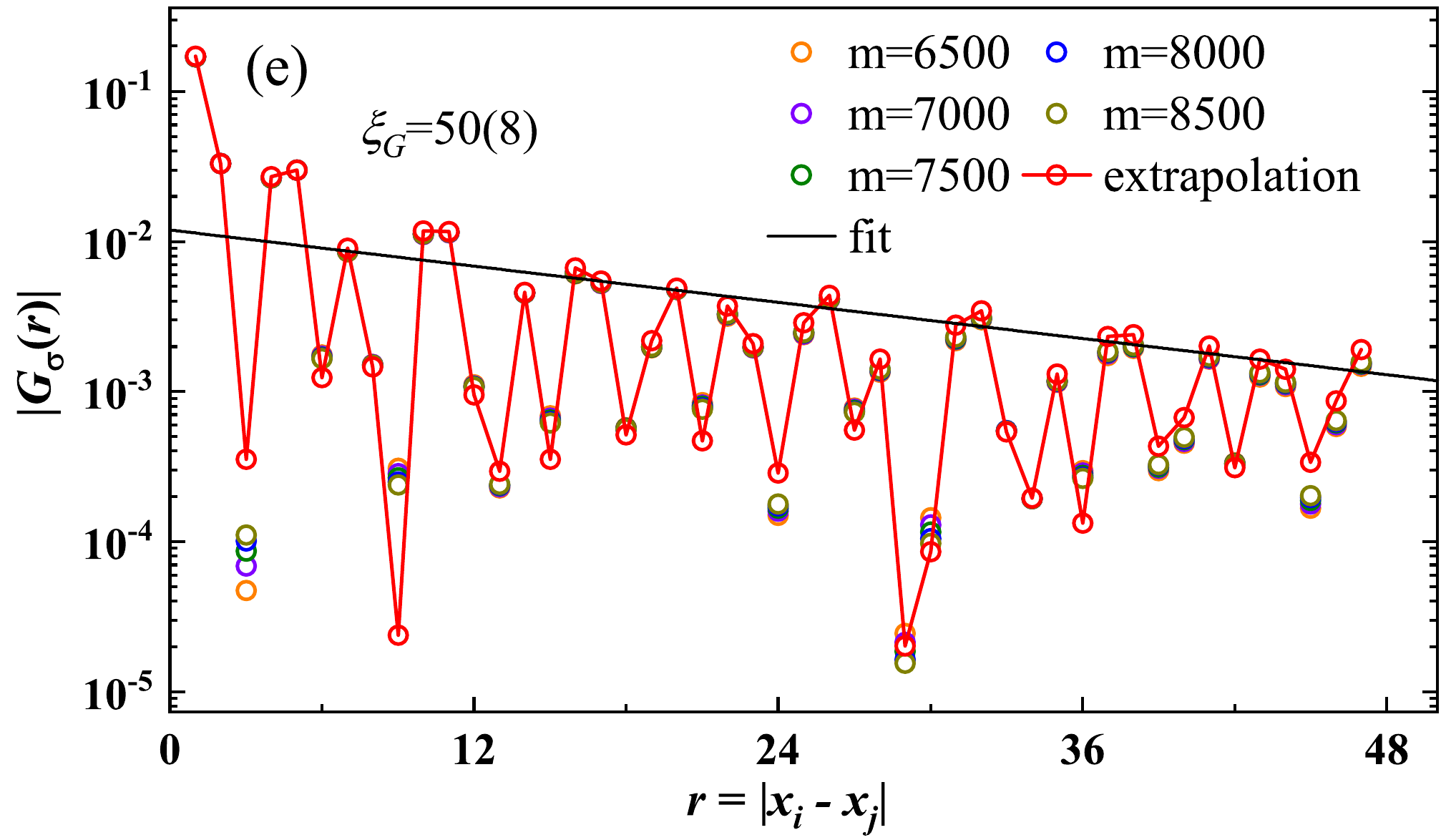}
    \includegraphics[width=0.4\textwidth]{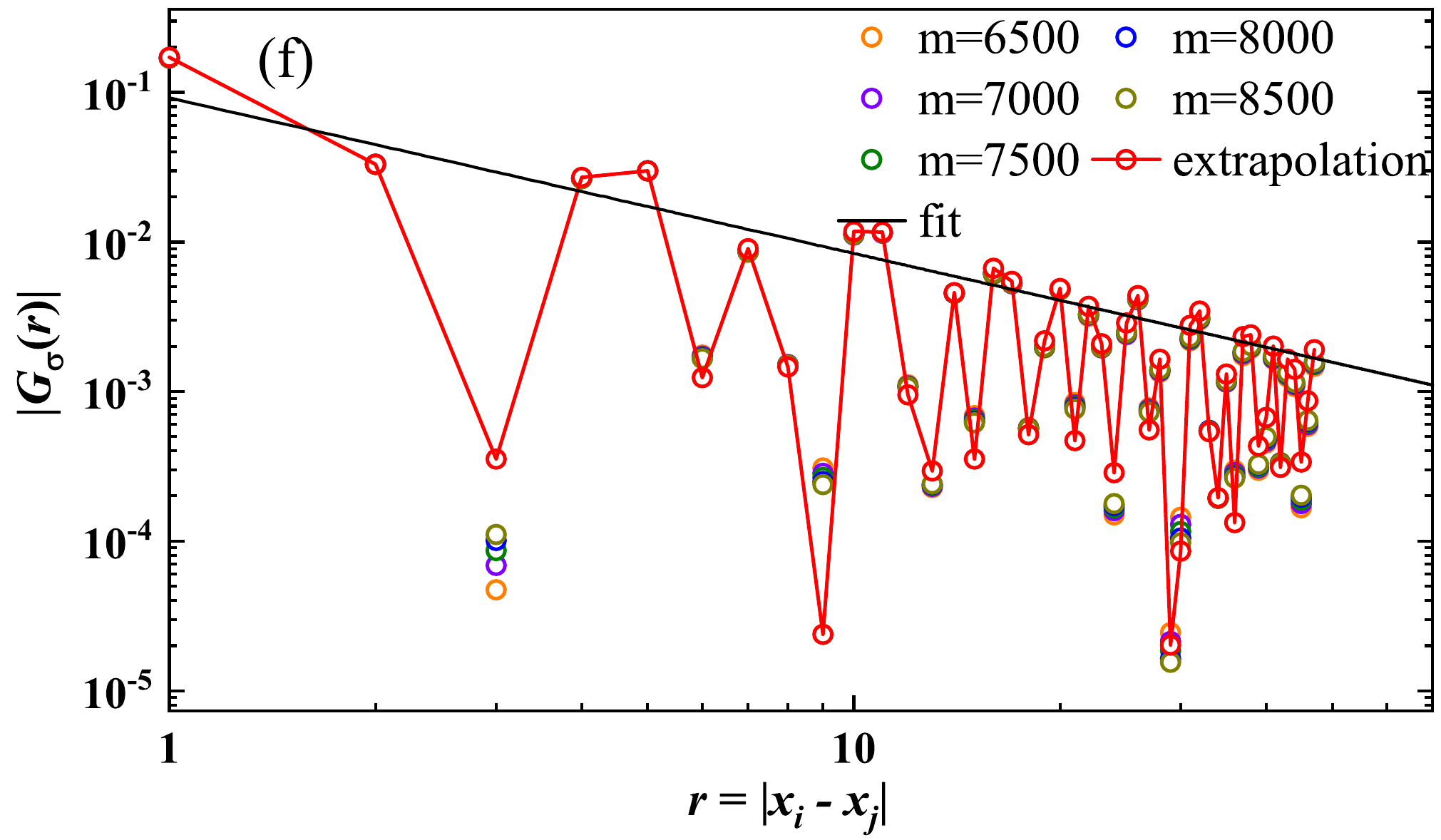}
    \caption{Pair-pair correlation, local magnetization and single-particle Green's function at $1/3$ doping. (a) $K_{sc}$ as a function of the reference bond. The length of the system is $L_x = 96$. (b) A typical fit of pair-pair correlation function. The reference bond is set at the 49th vertical bond. Absolute values of spin density (c, d) and single-particle Green's function (e, f). The results from finite kept state and the extrapolated with truncation errors results are shown. The solid lines in (c, e) and (d, f) denote exponential and power-law fits, respectively.For spin density, we find the data can be fitted with either exponential or power law. But the correlation length extracted from the power law fit is larger than the system size which indicates the decay of the local magnetization is likely to be power law. For single-particle excitation, we also find the decay behavior of $G_{\sigma}(r)$ can be fitted with either exponential or power law. Only peaked values are used in the fits.}
    \label{Onethird}
\end{figure*}

\begin{acknowledgments}
We thank useful discussions with Xingjie Han.
Y. Shen and M. P. Qin thank Weidong Luo for his generosity to provide computational resources for this work. M. P. Qin acknowledges the support from the National Key Research and Development Program of MOST of China (2022YFA1405400), the National Natural Science Foundation of China (Grant No. 12274290), the Innovation Program for Quantum Science and Technology (Grant No. 2021ZD0301900) and the sponsorship from Yangyang Development Fund. All the DMRG calculations are carried out with iTensor library\cite{itensor-r0.3}.
\end{acknowledgments}

\newpage
\bibliography{Hubladder}

\end{document}